\documentclass[twocolumn,showpacs,preprintnumbers,amsmath,amssymb,showkeys]{revtex4}
\usepackage{graphicx}
\usepackage{dcolumn}
\usepackage{bm}

\begin{document}

\title{
	Memory-function approach to the normal-state
	optical properties of the  Bechgaard salt (TMTSF)$_2$PF$_6$
	}
\author{
 	Ivan Kup\v{c}i\'{c}\footnote{
	E-mail address: kupcic@phy.hr
	(I. Kup\v{c}i\'{c})}
	}
\affiliation{ Department of Physics, 
	Faculty of Science, P.O.B. 331,  HR-10 002 Zagreb, Croatia
	}

\begin{abstract}
The gauge invariant, two-component optical conductivity model,
with a correlation gap structure related to the umklapp 
scattering processes, is 
applied to the quasi-one-dimensional electronic systems 
and compared to the recent measurements on  the 
Bechgaard salt (TMTSF)$_2$PF$_6$.
	The optical response of both the insulating and metallic state
is found for the half-filled conduction band, depending on the ratio
between the correlation energy scale $2 \Delta^0_2$ and
the transfer integral in the direction perpendicular to the conducting
chains, $t_{{\rm b}'}$.
	The estimated value $2 \Delta^0_2/t_{{\rm b}'}$ agrees reasonably well 
with the previous experimental and theoretical conclusions.
	Parallel to the chains the thermally activated conduction electrons in
the insulating state are found to exhibit an universal behaviour,
accounting for the observed single-particle optical
conductivity of the ordered ground state of charge-density-wave systems.
	The band parameters  and the related damping energies suitable to
the normal metallic state of (TMTSF)$_2$PF$_6$ are estimated  from the 
measured spectra.
	Not only the  spectral weights but also the  damping
energies clearly indicate an opening of the correlation gap in the charge 
excitation spectrum.
\end{abstract}



\noindent

\maketitle

\newpage

\section{Introduction}
%

The electrodynamics of  quasi-one-dimensional (Q1D) and two-dimensional 
(2D) electronic systems
has been the subject of  intensive experimental investigation
\cite{Bechgaard,Jacobsen,Dressel,Schwartz,Degiorgi2,Vescoli,Degiorgi,Uchida,Lupi,Venturini},
in part because various  correlation
and dimerization gaps in the charge excitation spectrum are observed recently 
in the measurements conducted on several high quality single crystals.
	Strong evidence for an unusual metallic state in the Bechgaard salts
(TMTSF)$_2X$, where $X =$ PF$_6$, AsF$_6$ or  ClO$_4$, has been found 
not only at temperatures close to the critical spin-density-wave (SDW) 
temperature, $T_{\rm SDW}$, but also at temperatures well above 
$T_{\rm SDW}$ \cite{Schwartz}.

The  anomalies  observed in (TMTSF)$_2$PF$_6$ in the latter temperature 
range are the focus of the present analysis (the data measured at 
$T \approx 20$~K and $T \approx 100$~K \cite{Dressel,Schwartz,Vescoli} 
will be considered here as the typical representatives of
the $T \approx T_{\rm SDW}$ and $T \gg T_{\rm SDW}$ spectra).
	The most pronounced anomalous features characterizing the spectra
measured for the electromagnetic fields polarized in the highly
conducting direction are as follows.
	(i) The spectral weight of the zero-frequency  peak is 
almost independent of
temperature, with the related effective number of conducting electrons 
$n^{\rm eff}_{{\rm intra},a} \approx n/100$ indicating the deconfined
(metallic) state of the electronic system ($n = 1/V_0 $ is  the hole 
concentration, with $V_0 $ being the primitive cell volume).
	(ii) There is a pronounced  deviation of the  zero-frequency
conductivity from  a simple Drude behaviour which points at substantial
frequency corrections in the ``intraband'' damping energy 
$\hbar \Gamma_{\rm intra}$.
	(iii) The maximum in the mid-infrared (MIR) optical conductivity 
shifts from $\omega_{\rm peak}/(2 \pi c) = 250 \; {\rm cm}^{-1}$ at 
$T = 20$ K to $\omega_{\rm peak}/(2 \pi c) = 800 \; {\rm cm}^{-1}$ 
at $T = 300$ K, which is approximately the temperature dependence expected for 
the ``interband ''damping energy $\hbar \Gamma_{\rm inter}$.
	(iv) This peak broadens with temperature exhibiting both a strong 
subgap conductivity (with 
$\hbar \Gamma_{\rm inter} \approx 30$~meV at room
temperatures, estimated in Ref.~\cite{Pedron}) and a non-universal
(temperature-dependent) power-law behaviour of the optical conductivity 
in the frequency range 
$ \omega \gg \omega_{\rm peak}$.

In order to discuss these anomalies
and the existing theories describing the electrodynamics of the 
strongly correlated  strictly 1D or Q1D systems 
\cite{Pedron,LRA,Gruner,Maki,Kivelson,Giamarchi,Mila,KupcicPB,Suzumura,Perroni},
we will consider here a simple, single-particle model of the conduction 
electrons in the presence of both two  gaps
and   external electromagnetic fields.
	It is already shown that such a model gives rise the exact solution
of the site-energy dimerization  problem (for example,
the anion ordering in (TMTSF)$_2$ClO$_4$) \cite{KupcicUP},
as well as a good approximation for the weak bond-energy dimerization
(and the related single-particle properties of the crystals with
the charge-density-wave (CDW) ground state) \cite{KupcicPB}.
	Here we will show that this model also provides a reasonable
description of the correlation effects related to the umklapp scattering
processes.
	Unlike the exact treatment of the umklapp  in the 
strictly 1D approaches based on the bosonization procedure
\cite{Giamarchi,Suzumura},
suitable to temperatures close to $T_{\rm SDW}$,
the present model describes the electrodynamics of the 
interacting electronic system at $T \gg T_{\rm SDW}$ 
in terms of  the effective fermions,
and, consequently, is the subject of easy comparison with similar
fermionic analyses  concerned with the room temperature
electrodynamic features \cite{Pedron,Mila}
or with other (anomalous) properties
of the normal metallic state of these salts 
\cite{Bechgaard,Voit,Chaikin,Gabovich}.

The article is organized as follows.
	In Section 2 and Appendix A  we present the two-component
optical conductivity for the Q1D electronic system with the 
dimerization/correlation gap in the single-particle excitation spectrum,
which is a simple extension  of the memory-function formalism
\cite{Gotze,Giamarchi,Forster,Plakida}.
	The confinement of conduction electrons to the highly conducting 
chains is discussed in Section 3, by considering the spectral weights
of the zero-frequency and MIR optical conductivity
as a function of the crossover parameter, i.e. of the ratio between 
the gap magnitude and the 
transfer integral in the direction perpendicular to the chains.
	The interplay among the small parameters of the two-component optical
model  is illustrated then for the insulating and metallic state 
of the half-filled band.
	The results are compared to the experimental spectra of
(TMTSF)$_2$PF$_6$.
	The concluding remarks are given in the last section.

\section{Theoretical model}
%

In this section, the memory-function  theory will be applied to determine 
the response of a simple Q1D  metallic system to the external 
electromagnetic fields.
	The electrons are assumed to be affected by  two
single-particle perturbations $\Delta_m  ({\bf k})$ with  the 
commensurate wave vectors ${\bf Q}_m = ( 2\pi /(ma), 0)$, 
$m = 1$ and 2 (where ${\bf Q}_{2} \approx 2{\bf k}_{\rm F}$).
	${\bf a} = a \hat{{\bf x}}$ and ${\bf b}' = b' \hat{{\bf y}}$ 
are the primitive vectors of the rectangular Bravais  lattice, and the $x$-axis 
is along the highly conducting direction.
	For the Bechgaard salts, $\Delta_1  ({\bf k})$ is the potential related 
to the bond-energy dimerization along the $x$-axis, and $\Delta_2  ({\bf k})$ 
is an effective perturbation which will model the opening of the pseudogap 
in the charge excitation spectrum caused by the umklapp scattering processes.
	However, it must be noted that the present consideration 
will be restricted to temperatures well 
above $T_{\rm SDW}$ ($T_{\rm SDW} \approx 12 $ K in 
(TMTSF)$_2$PF$_6$ \cite{Schwartz}), and, consequently, the nesting features 
of the Fermi 
surface responsible for the SDW instability of the electronic system 
will be disregarded.

In order to make  the generalization of the  memory-function approach more 
straightforward, the analysis is divided into three steps.
	For the crystallographic data characterizing the Bechgaard salts, 
the direct effects of $\Delta_1  ({\bf k})$ on the current-current correlation 
function are   found to be negligible~\cite{Pedron}. 
	This issue is briefly reconsidered in Section 2.1.
	In Section 2.2 the corresponding indirect effects are parametrized 
in terms of the  umklapp interaction $g_{1/2}$~\cite{Barisic},
and treated by means of the mean-field approximation.
 	The current-current correlation functions are determined then for 
the case of low impurity concentration.
	Finally, after ensuring that the resulting optical conductivity obeys 
the causality properties, the residual scattering processes are taken 
into account in Section 2.3.
	
   \begin{figure}[tb]
     \includegraphics[height=16pc,width=21pc]{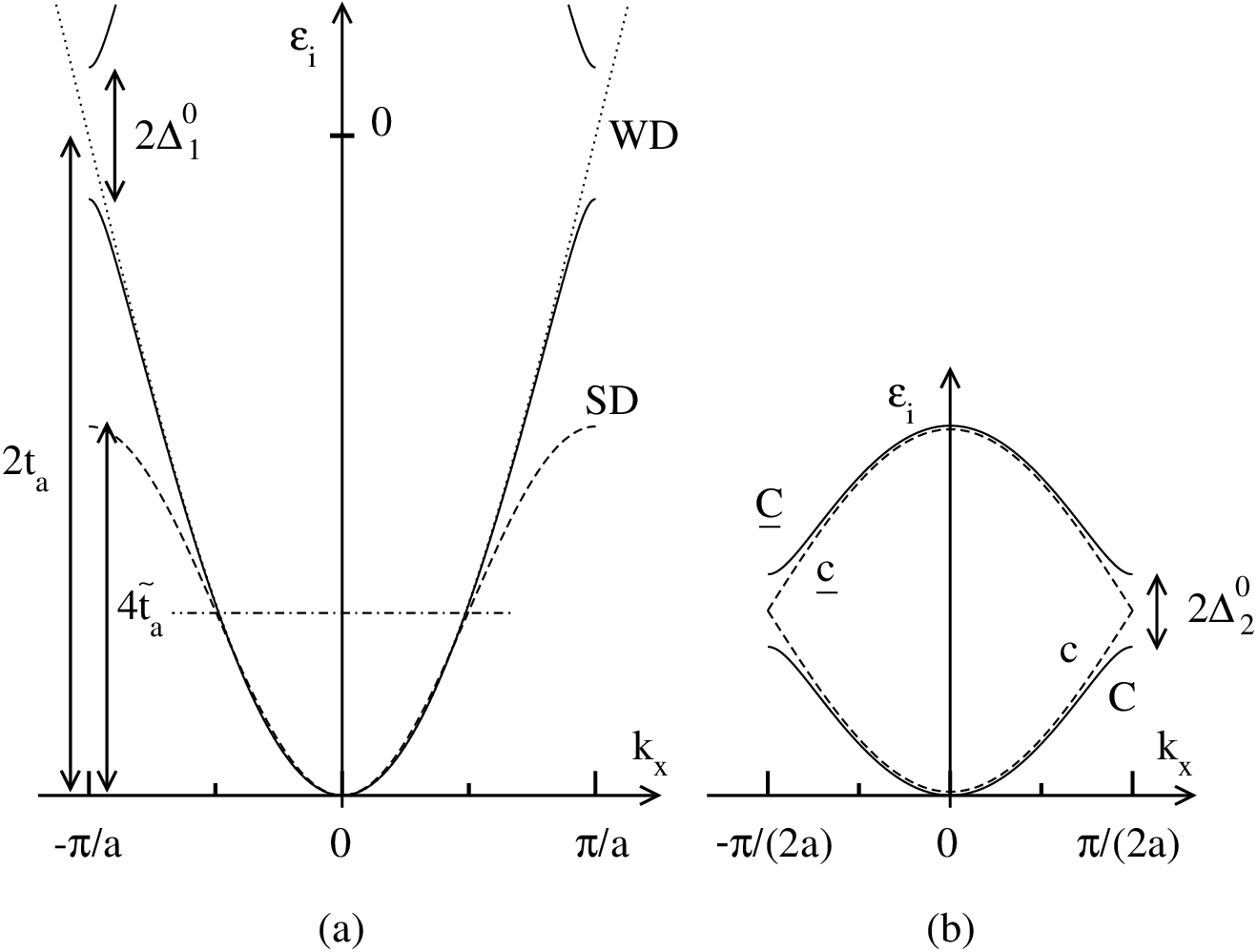}
    \caption{
    	The  hole dispersions $\varepsilon_i(k_x,  \pi/(2b'))$ (solid lines)
    as a function of the perturbations $\Delta_1  ({\bf k})$ (a)
    and $\Delta_2  ({\bf k})$ (b).
    	The $\Delta_1  ({\bf k}) \neq 0 $ dispersion  is contrasted to 
     the weak-dimerization (WD) and strong-dimerization (SD) dispersions
     defined in the text.
    	The dot-dashed line is the Fermi level in the quarter-filled 
    $\Delta_1  ({\bf k}) = 0$ band, and $\varepsilon_{\rm F} = 
    \varepsilon_0 (\pi/(2a), \pi/(2b')) - \varepsilon_0 (0, \pi/(2b'))$
    is the related Fermi energy.
    	The dispersions of the $\Delta_2  ({\bf k}) = 0$ and
    $\Delta_2  ({\bf k}) \neq 0$ SD models are shown in the reduced zone 
    representation; $c$, $\underline{c}$ and $C$, $\underline{C}$ are 
    the corresponding band indices.  
    }
     \end{figure}

\subsection{Bond-energy dimerization}
%

The bond-energy dimerization  along the highly conducting chain axis  
occurs in  all Bechgaard salts, and is usually described in terms of the 
dimensionless parameter $\alpha = (a_1 - a_2)/a$, with $a_1$ and $a_2$ being 
nearest-neighbor distances and $a = a_1 + a_2$.
	In (TMTSF)$_2$PF$_6$ the  characteristics
of the triclinic Bravais lattice are found to be
$a_1 - a_2 = 0.03$ \AA, $a = 7.297$ \AA, $b' = b \sin \gamma = 7.291$ {\AA }
and 
$V_0 = {\bf a} \cdot ({\bf b} \times {\bf c}) =
714.3$ \AA$^3$ \cite{Bechgaard,Thorup}.

In these compounds the conduction electrons are usually described 
using the quarter-filled hole band
$\varepsilon_0 ({\bf k}) = -2t_{\rm a} \cos \frac{1}{2} {\bf k} \cdot {\bf a} 
- 2t_{{\rm b}'} \cos {\bf k} \cdot {\bf b}'$, with the perturbation 
$\Delta _1 ({\bf k}) = 
{\rm i} g \alpha \sin \frac{1}{2}{\bf k} \cdot {\bf a} $.
	Here $t_{\rm a}$ and $t_{{\rm b}'}$ are the bond-energies for 
$\alpha =0$, and $g$ is the related electron--phonon coupling.
	After diagonalization the lower band becomes half-filled, with the 
dispersion $\varepsilon ({\bf k}) = (1/2)[\varepsilon_0 ({\bf k})
+ \varepsilon_0 ({\bf k} \pm {\bf Q}_1)]
- \sqrt{(1/4)[\varepsilon_0 ({\bf k}) - \varepsilon_0 ({\bf k} \pm {\bf Q}_1)]^2 
+ |\Delta _1 ({\bf k})|^2 }$, as illustrated in Fig.~1(a).

The spectral weight of the interband optical excitations relative to the 
total spectral weight
is given  approximately  by $[\Delta_1^0 / |\varepsilon ({\bf k}_{\rm F})|]^2$, 
with $\Delta_1^0 = g \alpha$ representing  the magnitude of the dimerization 
potential and $E_{{\rm g}1} \approx |2\varepsilon ({\bf k}_{\rm F})|$ 
is the related threshold energy
(see the difference between the total and intraband spectral 
weights for the case $\Delta ({\bf k}) \rightarrow \Delta_1 ({\bf k})$ in
Eqs. (\ref{eq28}), (\ref{eq21}), (\ref{eq29}) and (\ref{eqA6})).
	At least in  (TMTSF)$_2$PF$_6$ this ratio is  negligible
\cite{Pedron}. 
	Therefore, the direct effects of $\Delta _1 ({\bf k})$ on 
the optical conductivity will  not  affect the forthcoming analysis 
and we can write the bare hole dispersion in the form
$\varepsilon ({\bf k}) \approx \varepsilon_0 ({\bf k})$
(hereafter, this limit is referred to as the weak-dimerization (WD) limit,
see Fig.~1(a)).
	Consequently, the quantitative analysis of the spectral weights
in Sec. 2.3 will employ the WD dispersion.
	On the other hand, for the qualitative consideration of the 
low-frequency  spectra,  the strong-dimerization  (SD) dispersion 
is more adequate due to simple symmetry properties of both 
$\varepsilon ({\bf k} ) $ and the related  current and Raman vertex functions.

\subsection{Umklapp scattering processes in the mean-field approximation}
%
%
%
From the work of Emery et al.~\cite{Barisic} it is known that 
the electrodynamic features of the Bechgaard salts are affected by
the bond-energy dimerization dominantly through 
the umklapp interaction $g_{1/2} \propto U \Delta_1^0/\varepsilon_{\rm F} $.
	Their conclusions have been supported by the systematic investigation
of the optical conductivity in the MIR part of the spectra \cite{Vescoli}.
	It has been established that,
as $\alpha$ shifts to higher values, the energy 
$E_{{\rm peak}} = \hbar \omega _{{\rm peak}}$ becomes 
larger going from  $E_{{\rm peak}}  \approx 25$ meV 
in (TMTSF)$_2$PF$_6$ ($\alpha \approx 0.004$ \cite{Barisic}) to  
$E_{{\rm peak}} \approx 200$ meV in (TMTTF)$_2$PF$_6$ ($\alpha \approx 0.014 $)).
	Additional evidence for the crucial role of these scattering
processes was given by Favand and Mila \cite{Mila}, considering the MIR
optical features of the dimerized 1D Hubbard model.
	They have found that the dimerization potential, together with the 
strong Hubbard interaction $U$, results in a renormalized threshold energy 
$E_{{\rm g}2} \sim 2\Delta_1^0$, in addition to the bare threshold 
energy $E_{{\rm g}1}$  ($E_{{\rm g}2}  \ll E_{{\rm g}1}$).

	In order to study the influence of  these processes on the 
electrodynamics of the conduction electrons described by 
the (WD or SD)  hole dispersion $\varepsilon ({\bf k} ) $ 
at $T \gg T_{\rm SDW}$, we adopt now the mean-field approximation, with
the effective perturbation $\Delta_2 ({\bf k}) \propto g_{1/2}$.

\subsubsection{Model Hamiltonian}
%

As illustrated in Fig.~1(b),  the electrons are shown in the  zone 
representation $|k_{x}| < \pi/(2a)$, $|k_{y}| < \pi/b'$
(for simplicity, the 2D Brillouin zone of (TMTSF)$_2$PF$_6$ 
is approximated  by the rectangular zone).
	The band indices of $\Delta_2^0 = 0$  and $\Delta_2^0 \neq 0$
will be denoted by $l \in \{ c, \underline{c} \}$ and 
$L \in \{  C, \underline{C} \}$, respectively.
	The $\varepsilon_l ({\bf k})$ and $E_L ({\bf k})$ are the 
related  hole dispersions (obviously, 
$\varepsilon_c ({\bf k}) = \varepsilon ({\bf k})$ 
and $\varepsilon_{\underline{c}} ({\bf k}) = \varepsilon ({\bf k \pm Q}_2)$).
	For the sake of generality, the index $m = 2$ (in $\Delta_2 ({\bf k})$,
${\bf Q}_2$  and 
$E_{{\rm g}2} = E_{\underline{C} } ({\bf k}_{\rm F}) 
- E_{C}({\bf k}_{\rm F})$; $E_{{\rm g}2}$ is the MIR threshold energy) 
will be omitted and an arbitrary filling of the conduction band
will be assumed.
	The final results (Section~3) will be given for 
${\bf Q} \rightarrow {\bf Q}_2 \approx 2{\bf k}_{\rm F}$, with the 
index $m = 2$ explicitly written
\cite{comment2}.

The total  Hamiltonian reads as 
\begin{eqnarray}
  H &=& 
  H_0 + H'_1 + H'_2 + H^{\rm ext}.  
\label{eq1}
 \end{eqnarray}
	$H_0$ is the bare Hamiltonian.
 	$H'_1$  represents  the single-particle scattering processes (due to
impurities, phonons, CDW or SDW fluctuations) and $H^{\rm ext}$ is the 
coupling Hamiltonian which couples the  electromagnetic fields  to the current 
and Raman density fluctuations.
 	In the memory-function approach at temperatures well above $T_{\rm SDW}$
the residual two-particle scattering processes, represented here by $H'_2$,
play usually the marginal role and therefore will be neglected in the present 
section.

The bare Hamiltonian comprises two terms; 
the first one (${\cal H}_0$) includes the contributions which
 are diagonal in the $l {\bf k}$ representation, 
and the second one ($H'_0$) the off-diagonal contributions
coming from the single-particle potential 
$\Delta ({\bf k}) = e^{\mathrm{i} \phi_0} |\Delta  ({\bf k})|$:
\begin{eqnarray}
  H_0 &=& 
   \sum_{ll'{\bf k} \sigma}  H_0^{ll'} ({\bf k}) 
  l_{{\bf k} \sigma}^{\dagger} l'_{{\bf k} \sigma},
\label{eq2}
 \end{eqnarray}
$H_0^{cc} ({\bf k}) \equiv \varepsilon_c ({\bf k})$, 
$H_0^{\underline{c} \underline{c}} ({\bf k}) \equiv  
\varepsilon_{\underline{c}} ({\bf k}) $
and
$H_0^{c \underline{c}} ({\bf k} ) = \Delta ({\bf k}) $.
	The transformations
\begin{eqnarray}
 l_{{\bf k} \sigma}^{\dagger} &=& 
 \sum_L U_{\bf k} (l,L)  L_{{\bf k} \sigma}^{\dagger}, 
\label{eq3}
 \end{eqnarray}
lead to
\begin{eqnarray}
  H_0 &=& 
  \sum_{L {\bf k} \sigma}  E_L ({\bf k}) 
  L_{{\bf k} \sigma}^{\dagger} L_{{\bf k} \sigma},
\label{eq4}
 \end{eqnarray}	
with the dispersions
\begin{eqnarray}
E_{\underline{C}, C} ({\bf k}) &=& 
\frac{1}{2}[\varepsilon_{\underline{c}}({\bf k}) + \varepsilon_c({\bf k})]
\pm  \sqrt{\frac{1}{4}\varepsilon_{\underline{c}c}^2  ({\bf k}) 
+ |\Delta ({\bf k})|^2}.
 \nonumber \\
\label{eq5}\end{eqnarray}
	The transformation-matrix elements are given by
\begin{eqnarray} 
&& \left( \begin{array}{ll} 
  U_{\bf k} (c,C) & U_{\bf k} (c,\underline{C})  \\
  U_{\bf k} (\underline{c},C) & U_{\bf k} (\underline{c},\underline{C}) 
\end{array} \right)   
  \nonumber \\
&& 
\hspace{10mm}
= \left( \begin{array}{cc}
  \displaystyle \cos \frac{\varphi ({\bf k})}{2} &
  \displaystyle e^{-\mathrm{i}\phi_0} \sin \frac{\varphi ({\bf k})}{2}  
    \\  \displaystyle -e^{\mathrm{i}\phi_0} \sin \frac{\varphi ({\bf k})}{2} 
   &  \displaystyle \cos \frac{\varphi ({\bf k})}{2}  
\end{array} \right),  
\label{eq6}
 \end{eqnarray} 
where 
\begin{eqnarray}
\tan \varphi ({\bf k} ) &=&  \frac{2|\Delta ( {\bf k} )|}{
\varepsilon_{\underline{c}c} ({\bf k}) },
\label{eq7}\end{eqnarray} 
and $\varepsilon_{\underline{c}c}  ({\bf k}) = 
\varepsilon_{\underline{c}} ({\bf k}) - \varepsilon_c ({\bf k})$.

In the $L {\bf k}$ representation, one can write
\begin{eqnarray}
  H'_1  &=& 
  \sum_{LL'{\bf k} {\bf k}' \sigma}  
  V^{LL'} ({\bf k},{\bf k}')  
  L_{{\bf k} \sigma}^{\dagger} L'_{{\bf k}' \sigma}, 
  \label{eq8} \\
  H^{\rm ext}  &=& 
  \sum_{ LL'{\bf k}  \sigma}  \bigg\{ \bigg[
-\frac{1}{c} A_{\alpha} ({\bf q}) J_{\alpha}^{LL'} ({\bf k}) 
+ \delta_{L,L'}\frac{e^2}{2mc}
  \nonumber \\  & & 
\times A^2_{\alpha} ({\bf q}) 
\gamma ^{LL}_{\alpha \alpha} ( {\bf k};2) \bigg] 
  L_{{\bf k}+ {\bf q} \sigma}^{\dagger} L'_{{\bf k} \sigma} 
  + {\rm h. c.} \bigg\}.  
\label{eq9}
 \end{eqnarray} 
	The electromagnetic fields are described in terms of the transverse 
vector potential $A_{\alpha} ({\bf r})$ polarized in the $\alpha$ direction
($A_{\alpha} ({\bf q})$ and $A_{\alpha}^2 ({\bf q})$ are the Fourier transforms
of $A_{\alpha} ({\bf r})$ and $A_{\alpha}^2 ({\bf r})$, respectively),
while $J_{\alpha}^{LL'} ({\bf k}) $ and
$\gamma ^{LL}_{\alpha \alpha} ( {\bf k};2)$ are the related current 
and bare Raman vertex functions (see Appendix~A).

The memory-function approach will be applied now to the Hamiltonian
(\ref{eq1}).

\subsubsection{Current--current correlation functions}
%

The  retarded current--current correlation functions are defined 
by \cite{Gotze}
\begin{eqnarray}
\Pi_{\alpha \alpha}^{LL'} ({\bf q},z) &=& 
\frac{1}{\hbar V} 
\langle \langle \hat{J}_{\alpha}^{LL'} ({\bf q});
\hat{J}_{\alpha}^{L'L} (-{\bf q}) \rangle \rangle _z,
\label{eq10}
 \end{eqnarray}
with $\hat{J}_{\alpha}^{LL'} ({\bf q})$ representing 
the current density operator 
\begin{eqnarray}
\hat{J}_{\alpha}^{LL'} ({\bf q}) &=& \sum_{ {\bf k} \sigma}
\hat{J}_{\alpha}^{LL'} ({\bf k},{\bf k}_+), \nonumber \\
\hat{J}_{\alpha}^{LL'} ({\bf k},{\bf k}_+) &=& 
J_{\alpha}^{LL'} ({\bf k}) L_{{\bf k} \sigma}^{\dagger} 
L'_{{\bf k}+ {\bf q}  \sigma}.
\label{eq11}
 \end{eqnarray}
	Here $ {\bf k}_{+}$ is the abbreviation for ${\bf k}+ {\bf q}$,
$z =  \hbar \omega  + {\rm i} \eta$ and 
$\langle \langle \hat{A}; \hat{B} \rangle \rangle _z$ labels the retarded
correlation function of the density operators $\hat{A}$ and $\hat{B}$:
\begin{eqnarray}
\langle \langle \hat{A};\hat{B} \rangle \rangle _z &=& 
\int_{-\infty}^{\infty} {\rm d} t \;
e^{ -{\rm i} z t/ \hbar}
\langle \langle \hat{A};\hat{B} \rangle \rangle _t,
\nonumber \\
\langle \langle \hat{A};\hat{B} \rangle \rangle _t &=& 
- {\rm i} \Theta (t) \langle [ \hat{A}(t),\hat{B}] \rangle
\label{eq12}
 \end{eqnarray}
($\hat{A}(t)$ is the Heisenberg representation of  $\hat{A}$).

In order to determine $\Pi_{\alpha \alpha}^{LL'} ({\bf q},z)$,
which is now a rather complicated function of various scattering terms
($ H_0'$, $H_1'$ or $H_2'$), one usually starts with 
the equation of motion for the correlation function 
\begin{eqnarray}
{\cal D}_1^{LL'} ({\bf k},{\bf k}',{\bf q}, t) &=&
 \langle \langle \hat{J}_{\alpha}^{LL'} ({\bf k},{\bf k}_+);
\hat{J}_{\alpha}^{L'L} ({\bf k}'_+,{\bf k}') \rangle \rangle _t,
\nonumber \\
\label{eq13}
 \end{eqnarray}
which gives
\begin{eqnarray}
 && z_{LL'} ({\bf k})   {\cal D}_1^{LL'} ({\bf k},{\bf k}',{\bf q},z) 
\label{eq14} \\
 && \hspace{5mm} = \delta_{{\bf k}, {\bf k}'} \hbar 
 |J_{\alpha}^{LL'} ({\bf k})|^2  
\big[ f_L({\bf k}) -f_{L'}({\bf k}_+) 
\big] - \frac{1}{z_{LL'} ({\bf k}')} 
   \nonumber \\
&&   
\hspace{5mm} \times  \big[
{\cal D}_2 ^{LL'} ({\bf k},{\bf k}',{\bf q}, z_{LL'} ({\bf k}'))
-
{\cal D}_2 ^{LL'} ({\bf k},{\bf k}',{\bf q},0) \big]. 
 \nonumber 
\end{eqnarray}
	The first term on the right-hand side of this expression is relevant 
only for the interband processes, and consequently was absent in the 
previous memory-function analyses \cite{Gotze,Giamarchi,Forster,Plakida}.
	In this term the effects of $H_0'$ on the photon absorption/emission
are described using the exact diagonalization of  the single-particle
problem, Eqs.~(\ref{eq2})--(\ref{eq7}), with the momentum conservation in the 
interband current vertices fulfiled due to ${\bf Q}$ in $H_0'$.
	On the other hand, the second term defines the intraband 
relaxation function, treating  the  relaxation processes (proportional to 
$(H')^2$, $H' \approx H'_1 $) in the way similar to the Boltzmann 
equations \cite{Ziman,comment3}.

Here ${\cal D}_2 ^{LL'} ({\bf k},{\bf k}',{\bf q},z_{LL'} ({\bf k}')) $ 
is the force-force
correlation function \cite{Mahan,Gotze,Plakida} associated with the 
current-current correlation function 
${\cal D}_1^{LL'} ({\bf k},{\bf k}',{\bf q},z) $:
\begin{eqnarray}
&&{\cal D}_2 ^{LL'} ({\bf k},{\bf k}',{\bf q},z_{LL'} ({\bf k}'))  
\label{eq15}
 \\ \nonumber 
 &&  \hspace{10mm}  
= \langle \langle 
[\hat{J}_{\alpha}^{LL'} ({\bf k}, {\bf k}_+),H'];
[\hat{J}_{\alpha}^{L'L} ({\bf k}'_+,{\bf k}'),H'] \rangle \rangle _z.
\end{eqnarray}
	Furthermore, 
$z_{LL'} ({\bf k}) = z +E_{L} ({\bf k}) -  E_{L'} ({\bf k}_+) $ and
$f_L( {\bf k}) = [1 + e^{\beta [E_L({\bf k}) - \mu]}]^{-1}$ 
is the Fermi--Dirac distribution function.

\subsubsection{High-frequency limit}
It is rather a good approximation to retain in the interband 
relaxation function only  self-energy contributions \cite{KupcicUP}, 
show this term in the form 
$- {\rm i} \eta_{\rm inter} {\cal D}_1^{LL'} ({\bf k},{\bf k}',{\bf q},z)$,
neglect in Eq.~(\ref{eq14}) the terms  which are proportional to ${\bf q}$
and, finally, focus  attention on the simplest form of the 
scattering Hamiltonian where the  interactions 
$V^{CC} ({\bf k}, {\bf k}')$ are time-independent.

The leading contributions  in $H'$ to 
${\cal D}_1^{LL'} ({\bf k},{\bf k}',{\bf q},z)$ are given now by
\begin{eqnarray}
z{\cal D}_1^{CC} ({\bf k},{\bf k}',{\bf q},z) &=&  -\frac{1}{z} \big[
{\cal D}_2 ^{CC} ({\bf k},{\bf k}',{\bf q}, z)
\nonumber \\
&& \hspace{5mm}-
{\cal D}_2 ^{CC} ({\bf k},{\bf k}',{\bf q},0) \big], \nonumber \\
{\cal D}_1^{C\underline{C}} ({\bf k},{\bf k}',{\bf q},z) &=&
\delta_{{\bf k}, {\bf k}'} |J_{\alpha}^{C\underline{C}} ({\bf k})|^2 
{\cal D}_{C\underline{C}} ({\bf k}, {\bf k}, z), 
\nonumber \\
\label{eq16}
 \end{eqnarray}
with the high-frequency intra- and interband current-current correlation 
functions ($L L'= CC$ and $LL' = C\underline{C}, \underline{C}C$, respectively)
of the form
  \begin{eqnarray}
&&z \Pi_{\alpha \alpha}^{\rm intra, \infty} (z) = 
\frac{1}{V} \sum_{ {\bf k} {\bf k}' \sigma}  
\frac{\langle |V^{CC} ({\bf k}, {\bf k}')|^2 \rangle }{z}  
\big[
 J_{\alpha}^{CC}  ({\bf k}) 
\nonumber \\ 
 &&
  -  J_{\alpha}^{CC}  ({\bf k}')\big] ^2
  \frac{1}{\hbar} [{\cal D}_{CC} ({\bf k}, {\bf k}', z)
- {\cal D}_{CC} ({\bf k}, {\bf k}', 0)], 
\label{eq17}
  \\
&&\Pi_{\alpha \alpha}^{\rm inter, \infty} (z) = 
\frac{1}{V} \sum_{{\bf k}  \sigma}  
[J_{\alpha}^{C\underline{C}}  ({\bf k}) ]^2
\frac{1}{\hbar} [{\cal D}_{C\underline{C}} ({\bf k}, {\bf k}, z) 
\nonumber \\ 
&&
\hspace{20mm}
+ {\cal D}_{\underline{C}C} ({\bf k}, {\bf k}, z)] 
\label{eq18}
 \end{eqnarray}
(the label $\infty$ stands for the high-frequency limit
$\hbar \omega / \eta_{\rm intra} \gg 1$; $\eta_{\rm intra}$ 
is the intraband damping energy to be defined below).
	All physically relevant contributions are shown here in terms
of
\begin{eqnarray}
{\cal D}_{LL'}  ({\bf k}, {\bf k}',z) &=& 
\langle \langle 
 L_{{\bf k}\sigma}^{\dagger} L'_{{\bf k}' \sigma};
 L'^{\dagger}_{{\bf k}'\sigma} L_{{\bf k} \sigma} 
  \rangle \rangle _z
 \nonumber \\
  &=& \hbar \frac{f_L({\bf k}) - f_{L'}({\bf k}')}{
  z + E_L({\bf k}) - E_{L'}({\bf k}')}.
\label{eq19}
 \end{eqnarray}

For example, in the single-impurity-scattering approximation, one obtains
$\langle |V^{CC} ({\bf k}, {\bf k}')|^2 \rangle \approx 
N_i/N |V^{0} ({\bf k}, {\bf k}')|^2$;
$N_i$ is the presumably small number of impurities and
$V^{0} ({\bf k}, {\bf k}')$ is the scattering potential due to one 
impurity \cite{Mahan}.
	The generalization of Eq.~(\ref{eq17}) to the case where the  
interactions $V^{CC} ({\bf k}, {\bf k}')$ are time-dependent
or where the two-particle scattering term  $H'_2$ is taken into account 
explicitly is straightforward  and will not be considered here
(see, for example, Ref.~\cite{Gotze}).

\subsubsection{Low-frequency limit}
In the memory-function approach to the normal metallic state, 
the  low-frequency current--current 
correlation functions come on  collecting the most singular scattering 
events in powers of $(H')^2/\omega $.  
	The standard procedure \cite{Gotze} is based on the intraband 
memory (or relaxation) function which is defined by
\begin{eqnarray}
M_{\alpha} (z) &=& 
- \frac{ m z \Pi_{\alpha \alpha}^{{\rm intra}, \infty} (z)}{
e^2 n^{\rm eff}_{\rm intra, \alpha}}.
\label{eq20}
 \end{eqnarray} 
	$n^{\rm eff}_{\rm intra, \alpha}$
is the effective number of conduction electrons \cite{comment1,comment5}
\begin{eqnarray}
n_{{\rm \rm intra}, \alpha}^{\rm eff} &=& \frac{m}{e^2} \frac{1}{V} 
\sum_{ {\bf k} \sigma} 
[J^{CC}_{\alpha} ( {\bf k})]^2 
(-)\frac{\partial f_C ({\bf k})}{\partial E_C ({\bf k})} \nonumber \\
&\equiv& 
\frac{1}{V} \sum_{{\bf k} \sigma }
\gamma^{CC}_{\alpha \alpha} ( {\bf k})   f_C({\bf k}),
\label{eq21}
 \end{eqnarray} 
with $\gamma^{CC}_{\alpha \alpha} ( {\bf k})$ being the static Raman vertex,
Eq.~(\ref{eqA6}).

In the metallic systems with the anisotropic electron dispersion,
the numerator of Eq. (\ref{eq20}) has a complicated structure.
	Nevertheless, for $\hbar \omega \rightarrow 0$, a formal decoupling 
of two integrals in the expression 
(\ref{eq17}) is possible if  the average-inverse-relaxation-time 
approximation is applied.
	In this case, the inverse relaxation time 
\begin{eqnarray}
\frac{\hbar }{\tau_{\rm intra}  ({\bf k})} & \approx&   
 \sum_{{\bf k}'}  \langle |V ^{CC}({\bf k}, {\bf k}')|^2 \rangle 
\big[ 1 -  J_{\alpha}^{CC}  ({\bf k}')/J_{\alpha}^{CC}  ({\bf k}) 
\big] 
\nonumber \\
& &
  \times 2 \pi \delta [E_C ({\bf k}') - \mu ]
\label{eq22}
 \end{eqnarray}
is averaged over the Fermi surface, resulting in the intraband damping energy  
\begin{eqnarray}
  \eta_{\rm intra} & =& {\rm  Im} \{ M_{\alpha} ( 0 + {\rm i} \eta ) \} = 
  \langle \frac{\hbar}{\tau_{\rm intra} ({\bf k})}  \rangle  .
\label{eq23}
 \end{eqnarray}
	Finally, the resulting intra- and interband current-current 
correlation functions read as
\begin{eqnarray}
\Pi_{\alpha \alpha}^{\rm intra} (\hbar \omega) & \approx &
-\frac{e^2 n^{\rm eff}_{\rm intra, \alpha}} {m} 
\frac{{\rm i}\eta_{\rm intra}}{\hbar \omega + {\rm i} \eta_{\rm intra}}, 
\nonumber \\ 
\Pi_{\alpha \alpha}^{\rm inter} (\hbar \omega) & \approx &
 \Pi_{\alpha \alpha}^{{\rm inter}, \infty} 
 (\hbar \omega + {\rm i} \eta_{\rm inter} ). 
\label{eq24}
 \end{eqnarray}
	A more rigorous treatment of the intraband relaxation processes should
include the explicit  numerical calculation of two coupled integrals in the 
expression (\ref{eq17}); however, this is beyond the scope 
of the present work.

\subsection{Optical conductivity}
Due to causality, the structure of the multi-component optical conductivity 
comes on combining the expressions for the current--current correlation 
functions obtained for $\eta_i \rightarrow 0$  with the corresponding 
diamagnetic contributions, and  replacing $\eta_i$  with the 
frequency-independent damping energies $\hbar \Gamma_{i}$ (representing the
single-impurity  relaxations for the usual impurity concentrations as well
as the other  relaxation processes not considered in the 
$\eta_i \rightarrow 0$ model)~\cite{Mahan,KupcicPB}.
	Depending on the ratio among the band parameters,
this procedure will result in different  optical-conductivity expressions.
	The two-component optical conductivity 
(where $2\Delta^0/E_{\rm g} \approx 1$)	
and the generalized Drude formula ($2\Delta^0/E_{\rm g} \ll 1$)
are the most interesting cases.

As a first approximation it seems  natural to interpret the 
optical conductivity of (TMTSF)$_2$PF$_6$  in terms of the two-component 
model with the correlation energy scale $2\Delta^0$ comparable to 
the threshold energy $E_g $, and with both $\hbar \Gamma_{\rm intra}$ and
$\hbar \Gamma_{\rm inter}$  independent of frequency. 
	The total optical conductivity $\sigma_{{\alpha}}^{\rm total} (\omega)$ 
is  the sum of the intra- and interband  contributions, which are given by
the gauge-invariant expressions
\cite{Pedron,KupcicPB,Pines}
\begin{eqnarray}
\sigma_{{\alpha}}^{\rm intra} (\omega) &=& \frac{\mathrm{i}}{\omega}
\frac{e^2 n_{{\rm intra},{\alpha}}^{\rm eff}}{m} 
  \frac{ \hbar \omega}{\hbar \omega + \mathrm{i}\hbar \Gamma_{\rm intra}}
  ,  \label{eq25} \\ 
 \nonumber
\sigma_{{\alpha}}^{\rm inter} (\omega) &=& \frac{\mathrm{i}}{\omega}
\frac{1}{V} \sum_{{\bf k} \sigma} \frac{(\hbar \omega)^2
|J_{{\alpha}}^{\underline{C}C} ( {\bf k})|^2  
}{E_{\underline{C}C} ^2 (  {\bf k}) }
 \\ \nonumber
  &&\times 
\bigg\{  \frac{  f_C({\bf k}) - f_{\underline{C}}({\bf k})  
}{\hbar  \omega  -  E_{\underline{C}C} (  {\bf k})    + 
 \mathrm{i}  \hbar \Gamma_{\rm inter} } 
\\ &&
 +
\frac{  f_{\underline{C}}({\bf k}) - f_C({\bf k}) 
}{\hbar  \omega + E_{\underline{C}C} ( {\bf k})  + 
 \mathrm{i} \hbar \Gamma_{\rm inter} } 
\bigg\},
\label{eq26} 
 \end{eqnarray}
$E_{\underline{C}C} ( {\bf k}) = E_{\underline{C}} ( {\bf k}) 
- E_{C} ( {\bf k})$.

Since the residual scattering processes (corresponding primarily to the
frequency-dependent contributions in $H_1'$, or to the two-particle 
term $H_2'$) become increasingly pronounced at low frequencies, 
the intraband contribution (\ref{eq25}) should
in principle be replaced by the generalized Drude formula
\begin{eqnarray}
\sigma_{{\alpha}}^{\rm intra} (\omega) &=& 
  \frac{ 
\mathrm{i} \hbar e^2 n_{{\rm intra},{\alpha}}^{\rm eff}/m }{
 \hbar \omega (m (\omega)/m)  + 
  {\rm i} {\rm Im} \{ M_{\alpha} (\hbar \omega + {\rm i} \eta) \}}.
\label{eq27} 
 \end{eqnarray}
	The total optical conductivity is now the sum of the expressions
(\ref{eq27})  and (\ref{eq26}). 
	Here $m (\omega)$ results from the Kramers--Kroning relations,
with $H' = H'_1 + H'_2$ in Eq. (\ref{eq20}).
	Similarly, in the limit 
$2\Delta^0/E_{\rm g} \ll 1$,
one obtains the generalized Drude expression 
$\sigma_{{\alpha}}^{\rm total } (\omega) 
\approx \sigma_{{\alpha}}^{\rm intra} (\omega)$ with 
$H' = H'_0 + H'_1 + H'_2$ and $\sigma_{{\alpha}}^{\rm intra} (\omega)$
given by Eq. (\ref{eq27}).

	It should be noted that the direct integration of the real 
part of the expressions (\ref{eq25}) and (\ref{eq26}) gives rise to the 
intraband, interband  and total  spectral weights 
of the form \cite{KupcicPB,KupcicPC}
\begin{eqnarray}
 \frac{1}{2}\Omega^2_{{i},{\alpha}} = 
  \frac{m_{\rm aa}}{m} V_0 n^{\rm eff}_{{i},{\alpha}}  
 \frac{m}{m_{\rm aa}} 
 \frac{1}{2}\Omega^2_{0}, 
\label{eq28} 
 \end{eqnarray}
where $i \in \{ {\rm intra}, {\rm inter}, {\rm total} \}$.
	The effective total number of electrons is given by
\begin{eqnarray}
n^{\rm eff}_{{\rm total},{\alpha}} &=& \frac{1}{V} \sum_{{\bf k} \sigma }
\gamma^{CC}_{\alpha \alpha} ( {\bf k};2)   f_C ({\bf k})  ,
\label{eq29} 
 \end{eqnarray}
$n^{\rm eff}_{{\rm inter},{\alpha}} = n^{\rm eff}_{{\rm total},{\alpha}}
- n^{\rm eff}_{{\rm intra},{\alpha}}$ and $n^{\rm eff}_{{\rm intra},{\alpha}}$
comes from Eq.~(\ref{eq21}).
	Furthermore,  $ \Omega_0 = \sqrt{4 \pi e^2/(mV_0)}$ is an 
auxiliary frequency scale which represents the bare plasma frequency 
attributed to the free holes with the concentration  $n = 1/V_0$, 
$m_{\rm aa} = 2 \hbar^2/(t_{\rm a} a^2)$  is the mass parameter
and the $(m_{\rm aa}/m) V_0 n^{\rm eff}_{{i},{\alpha}}$ are the effective 
numbers shown in a convenient dimensionless form. 

  \begin{figure}[tb]
     \includegraphics[height=17pc,width=18pc]{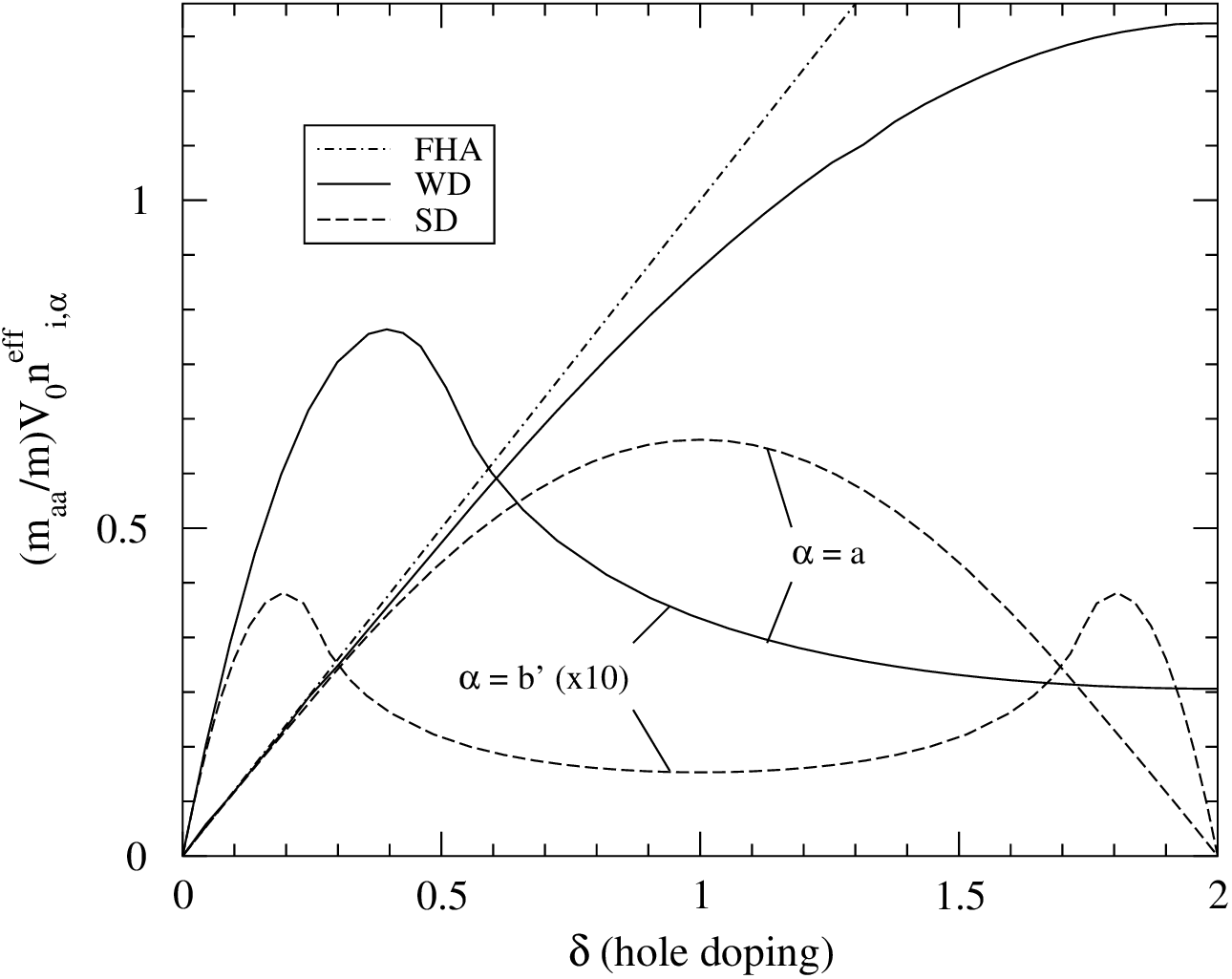}
     \caption{
     	The development of the effective number of conduction electrons 
     with doping for the $\Delta^0 = 0$ WD and SD models.
     	The dot-dashed line is the prediction of the free-hole approximation
     (FHA).
     $t_{\rm a} =   0.25$ eV, $t_{{\rm b}'} = 25$ meV (WD case) and
     $\tilde{t}_{\rm a} = 0.125$ eV, $t_{{\rm b}'} = 12.5$ meV (SD case).
     	For clarity, the $\alpha = b'$ effective numbers are multiplied by 
     a factor of 10.	 
    }
     \end{figure}

Before turning to the detailed numerical calculation, let us consider the 
total spectral weights and remember the 
estimates of the bond-energies in (TMTSF)$_2$PF$_6$
\cite{Jacobsen,Pedron}.
		It has been established that the measured plasma energy 
$\hbar \Omega_{{\rm total , a}}$ (1.1--1.4~{\rm eV} 
\cite{Bechgaard,Jacobsen,Dressel,Schwartz})
is close to the energy $\hbar \Omega_0  \approx 1.385$ eV
(or $ \Omega_0/(2 \pi c) \approx 1.1 \cdot 10^4$ cm$^{-1}$).
	Using the $\Delta^0 = 0$ WD model \cite{Jacobsen}  this observation 
has led to the conclusion that the bond-energy 
$t_{\rm a} \approx t_{\rm a} ^0$, where
$t_{\rm a}^0 \approx 0.285$ eV is the characteristic bond-energy for 
which the mass parameter $m_{\rm aa} $ is equal to the bare electron/hole 
mass $m$.
	Similarly, for $\delta \approx 1$ (corresponding to the open Fermi
surface), one 
obtains \cite{comment4}
\begin{eqnarray}
n^{\rm eff}_{{\rm total,b}'}  &\approx& 
\beta \frac{ t_{{\rm b}'}^2}{ t_{\rm a}^2} 
n^{\rm eff}_{{\rm total,a}},
\label{eq30}
 \end{eqnarray}
with $\beta \approx 4$ for the $\Delta^0 = 0$ WD case (see Fig. 2).
	For $t_{{\rm b}'} = 0.1  t_{\rm a}$, 
this has resulted in the bare plasma frequency
$\Omega_{{\rm total, b}'}  \approx 0.2  \Omega_{{\rm total, a}}$
which agrees reasonably well  with  measured values 
\cite{Jacobsen,Dressel,Schwartz}.

Due to the scattering term $H'_0$, for ${\bf Q} \approx 2{\bf k}_{\rm F}$ 
the quantities $n_{{\rm intra},{\alpha}}^{\rm eff}$ and 
$\Pi_{\alpha \alpha}^{{\rm intra}, \infty} (z)$ 
are strongly reduced with
respect to their values in the free-hole approximation.
	In addition, the analytical decoupling of two integrals in 
$\Pi_{\alpha \alpha}^{{\rm intra}, \infty} (z)$ can be
achieved only using the average-inverse-relaxation-time approximation
($-\hbar \omega {\rm Im} \{ \Pi_{\alpha \alpha}^{{\rm intra}, \infty} 
(\hbar \omega + {\rm i} \eta) \}
\approx  \hbar \Gamma_{\rm intra} 
e^2  n^{\rm eff}_{{\rm intra},{\alpha}}/m$). 
	For this reason, from here on we will focus the attention on the 
two-component model (\ref{eq25})--(\ref{eq26}) only, with the damping energies 
independent of frequency and with $\Delta ({\bf k}) = \Delta^0$.

Leaving the factors $m/m_{\rm aa}$ and $\Omega^2_{0}/2$ in
Eq.~(\ref{eq28}) off, the total  spectral weight is calculated now as 
a function of the doping level for the bond-energies 
$t_{\rm a} =   0.25$ eV, $t_{{\rm b}'} = 25$ meV (WD case) and
$\tilde{t}_{\rm a} = 0.125$ eV, $t_{{\rm b}'} = 12.5$ meV (SD case) 
and for the photon polarizations $\alpha = a$ and $\alpha =  b'$, 
and shown in Fig. 2.
	Although the SD model, when fitted to the measured total spectral 
weights,  will result in the slightly different values of the bond-energies, 
on the qualitative side (for not too large $\Delta^0/(2\tilde{t}_{\rm a})$
and $\hbar \omega/(2\tilde{t}_{\rm a})$)
there is no essential difference between these two models in explaining
the structure of the $\Delta^0 \neq 0$ optical conductivity.
	In this respect, we continue the analysis considering  the 
SD model, and assuming that 
$\hbar \omega, k_{\rm B} T, \Delta^0, t_{{\rm b}'}, \hbar \Gamma_i  
\ll  2\tilde{t}_{\rm a}$.
	It is important to notice that the model (\ref{eq25})--(\ref{eq26}) 
for $t_{{\rm b}'} \rightarrow 0$ is essentially the same as the strictly 1D
spinless-fermion optical model of Pedron et al. \cite{Pedron},
and, not surprisingly, the present estimates of $t_{\rm a}$, $\Delta^0$
and $\hbar \Gamma_i $ will be close to their estimated values.
	However,  there is a pronounced qualitative difference
between these two optical models regarding the Fermi energy 
and wave vector positions (${\bf Q} \approx 2{\bf k}_{\rm F}$ here,
contrary to ${\bf Q} \approx 4{\bf k}_{\rm F}$ in the spinless-fermion
model).

\section{Comparison with experiments}
%

The  optical anomalies of (TMTSF)$_2$PF$_6$  mentioned in Section 1 
will be analyzed now in the framework of 
the Q1D fermionic model (\ref{eq25})--(\ref{eq26}).
	The competition between the small energy scales of this model, 
which results in the structures in the optical conductivity similar to that 
observed in experiments, is the focus of the present discussion.
	We consider first a few general features which emerge as
a result of this competition.
	Then the model parameters suitable to the $T = 100$~K spectra of
(TMTSF)$_2$PF$_6$ will be discussed in more detail.
	In the rest of the text the full notation is used with the index 
$m=2$, in $\Delta_2^0 $ and $E_{{\rm g}2}$,  written explicitly.

\subsection{Competition between $t_{{\rm b}'}$ and $\Delta^0_2$}
%

  \begin{figure}[tb]
     \includegraphics[height=18pc,width=18pc]{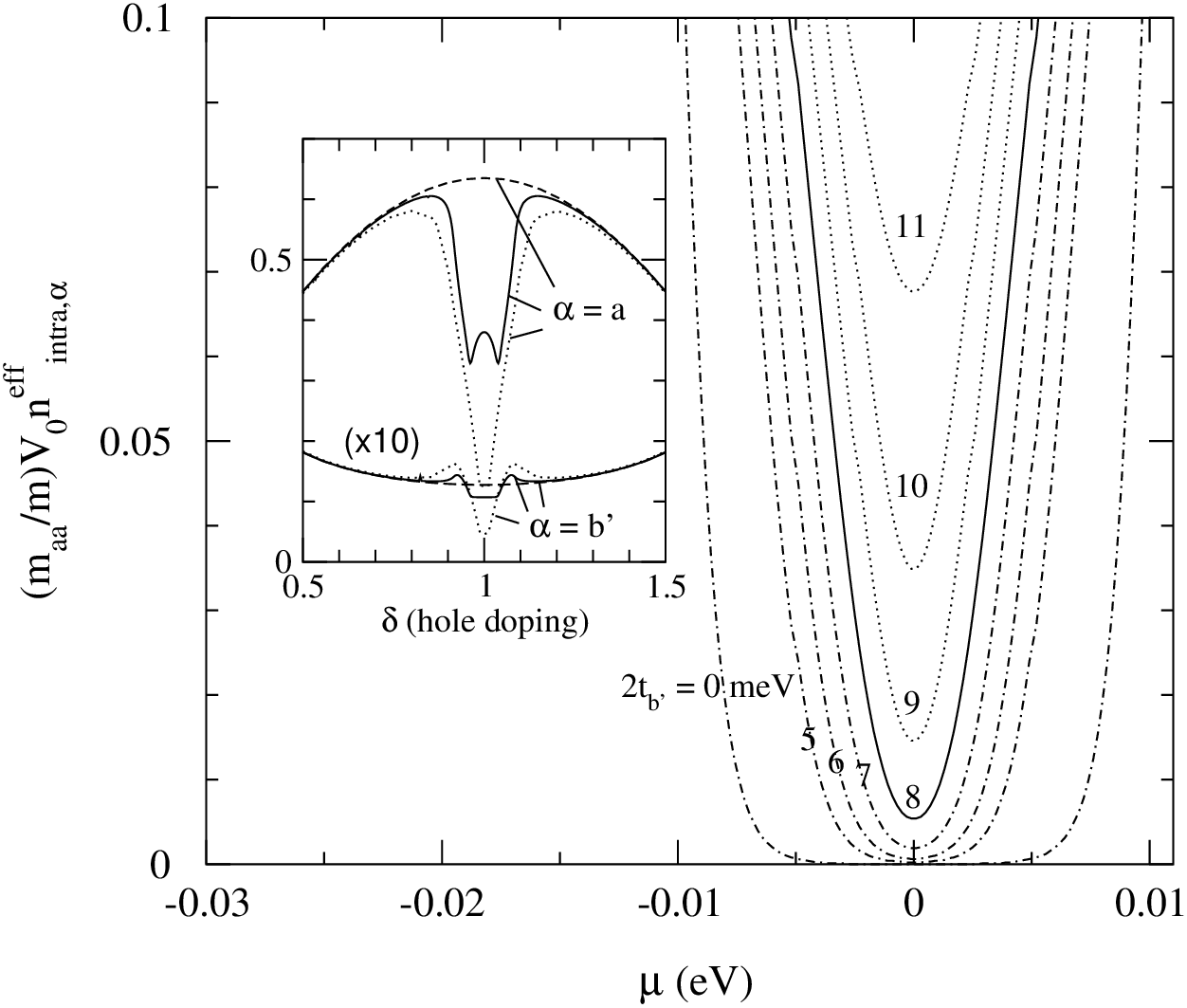}
    \caption{
    Main figure: The $\alpha = a$ intraband spectral weight  as a function 
    of the chemical potential $\mu$ for different values of $2t_{{\rm b}'}$
    ($2t_{{\rm b}'} = 0$ and $5-11$ meV, 
    $2\tilde{t}_{\rm a } = 0.25$ eV, $\Delta^0_2 = 10$ meV and $T = 10$ K). 
    Inset of figure: The dependence of $n_{{\rm intra},\alpha}^{\rm eff}$ 
    on $\delta$ for the $\Delta^0_2 = 0$ (dashed line), 
    $\Delta^0_2 = 10$ meV  (solid line) and
    $\Delta^0_2 = 20$ meV  (dotted line) SD models, 
    $2\tilde{t}_{\rm a } = 0.25$ eV, $2t_{{\rm b}'} = 25$ meV and 
    $T = 10$ K.
    The $\alpha = b'$ spectral weight is again multiplied by 10.
        }
     \end{figure}
     
The $\alpha = a$ conductivity sum rule of the present two-component model, 
$\Omega^2_{{\rm total}, {\rm a}} = \Omega^2_{{\rm intra}, {\rm a}}
+ \Omega^2_{{\rm inter}, {\rm a}}$,  does not depend on the
damping energies $\hbar \Gamma_i$.
	The main consequences of the interplay between $t_{{\rm b}'}$ 
and $\Delta^0_2$ can be thus seen from Fig. 3, where the effective numbers 
$n^{\rm eff}_{{\rm intra}, \alpha}$ are displayed as a function of the crossover 
parameter  $2\Delta^0_2 / t_{{\rm b}'}$ at a relatively small 
temperature $T = 10$~K.
	For $t_{{\rm b}'} \rightarrow 0$ but $\Delta^0_2$ finite, 
the electrons will be confined on individual chains, as found in the 
insulating (TMTTF)$_2X$ salts, where $X =$ PF$_6$ or Br \cite{Vescoli}.
	The  effective number $n^{\rm eff}_{{\rm intra},{\rm a}} \approx n/100$ 
required to give the metallic behaviour observed in (TMTSF)$_2$PF$_6$ 
is obtained for  $2\Delta^0_2 / t_{{\rm b}'} \approx 5$ (main figure,
$T = 10 $ K).
	This value  is approximately twice the critical value 
$E_{{\rm g}2}  / t_{{\rm b}'} =$~1.8--2.3 
for the confinement
(i.e. for the metal-to-insulator phase transition)
obtained by the renormalization group approach for the problem
of two coupled chains \cite{Suzumura}.
	The value of $2\Delta^0_2 / t_{{\rm b}'}$ decreases with decreasing 
temperature, resulting in   
$2\Delta^0_2 / t_{{\rm b}'} \approx 4$  at $T = 2 $ K. 
	Another reason for this discrepancy should be the fact that the ratio
$\tilde{t}_{\rm a} / t_{{\rm b}'} $  plays important role as well; namely, for
$\tilde{t}_{\rm a}$ fixed, the increase of $t_{{\rm b}'}$  transforms the Q1D
electronic system into an anisotropic 2D system, where the dimerization along
the  $x$-axis becomes less important.
	As   illustrated in the inset of the figure, for 
$2\Delta^0_2 / t_{{\rm b}'}$ small enough, the gap opens only on a part 
of the Fermi surface resulting for $n^{\rm eff}_{\rm intra, a}$ in a 
local maximum at $\delta = 1$.
	On the contrary, for $2\Delta^0_2 / t_{{\rm b}'} > 2.5$ and $T = 10$ K, 
the correlation gap develops on the complete Fermi surface, with 
the bare group velocity $J^{cc}_{\rm a} ( {\bf k})/e$ in the effective 
number of conduction electrons
replaced by the ``renormalized'' one,
$J^{CC}_{\rm a} ( {\bf k})/e = \cos \varphi ({\bf k}) 
J^{cc}_{\rm a} ( {\bf k})/e$, with approximately 
$\cos \varphi ({\bf k}) \approx \cos \varphi ({\bf k}_{\rm F} )$.

The number $n^{\rm eff}_{{\rm intra},{\rm b}'}$ is affected by 
the correlations in the same way as $n^{\rm eff}_{{\rm intra},{\rm a}}$.
	While the rest of the $\alpha = a$ total spectral weight, 
i.e. the difference between the dashed and solid (dotted) line in the inset 
of Fig.~3, reappears through the coherent  excitations across the 
correlation gap $E_{{\rm g}2} $, for $\alpha =  b'$ there is no 
counterpart of these coherent finite-frequency excitations.
	Therefore, to fulfil the $\alpha =  b'$ conductivity sum rule,
one has to go beyond the  optical model 
(\ref{eq25})--(\ref{eq26}), and to
take also incoherent optical processes in the ${\bf b}'$ direction 
into account \cite{Jerome}.
	Obviously, this issue is  of a general importance for the complete
quantitative description of the optical conductivity of the Bechgaard salts, and 
should be in relation with the disappearance of the
plasma edge in the $\alpha =  b'$ reflectivity spectra of
the (TMTTF)$_2X$ salts \cite{Vescoli}.

\subsection{Relation with the CDW systems and the high-$T_c$ cuprates}
Besides the dramatic dependence on the crossover parameter 
$2\Delta^0_2 / t_{{\rm b}'}$, the optical model (\ref{eq25})--(\ref{eq26})
exhibits also various temperature-dependent features, due to 
the competition between the energies 
$k_{\rm B} T$, $\Delta^0_2 (T)$ and $ \hbar \Gamma_{\rm inter} (T)$.
	Some of these temperature effects are at first sight similar to the
anomalies (i)--(iv) mentioned in Section 1.
	Interestingly, they are more directly related to different MIR optical 
features observed in various other Q1D and 2D systems.
	To compare briefly the predictions of the model with the 
MIR spectra measured in the CDW systems in the ordered ground state
\cite{Degiorgi}, 
as well as in the normal metallic state of the underdoped high-$T_c$ cuprates
\cite{Uchida,Lupi,Venturini}, 
we apply it now to the insulating $\delta = 1$, $t_{{\rm b}'} \rightarrow 0$ 
case.

The real part of the intraband conductivity has the usual form
${\rm Re} \{ \sigma_{\rm a}^{\rm intra} (\omega) \} = \sigma_{\rm a}^{\rm DC} 
/(1 + (\omega /\Gamma_{\rm intra})^2)$.
	$\sigma_{\rm a}^{\rm DC} = \Omega^2_{0} V_0 
n^{\rm eff}_{{\rm intra},{\rm a}} /(4 \pi  \Gamma_{\rm intra})$ 
is the DC conductivity, which vanishes at $T = 0$ K.
	The temperature dependence of $\sigma_{\rm a}^{\rm DC}$ 
is governed by the thermal activation of the conduction electrons,	
according to the expression
 \begin{eqnarray}
n^{\rm eff}_{{\rm intra},{\rm a}} &=& \frac{1}{V} \sum_{{\bf k} \sigma } 
\gamma^{CC}_{\rm aa} ( {\bf k}) [ f_C ({\bf k} )- f_{\underline{C}} ({\bf k}) ].
\label{eq31} \end{eqnarray}
	This temperature dependence is accompanied by the decrease 
of the interband spectral weight, as  seen from 
 \begin{eqnarray}
 {\rm Re} \{ \sigma_{{\rm a}}^{\rm inter} (\omega) \} &=& \frac{1}{\omega}
\frac{1}{V} \sum_{{\bf k} \sigma} \frac{(\hbar \omega)^2
|J_{{\rm a}}^{C\underline{C}} ( {\bf k})|^2  
}{E_{\underline{C}C} ^2 (  {\bf k}) } 
 \nonumber \\
 &&\times 
  \frac{4 \hbar \omega E_{\underline{C}C} (  {\bf k})}{[\hbar  \omega  
  +  E_{\underline{C}C} (  {\bf k})]^2    + 
   \hbar \Gamma_{\rm inter}^2 }
 \nonumber \\
 &&\times 
  \frac{ \hbar \Gamma_{\rm inter} 
  [ f_C ({\bf k} )- f_{\underline{C}} ({\bf k}) ]  
}{[\hbar  \omega  -  E_{\underline{C}C} (  {\bf k})]^2    + 
   \hbar \Gamma_{\rm inter}^2 }, 
\label{eq32} \end{eqnarray}
where the same temperature factor can be recognized
($f_C ({\bf k} )- f_{\underline{C}} ({\bf k}) \approx
\tanh [E_{\underline{C}} ({\bf k})/(2k_{\rm B}T)]$ for 
$\mu = 0$).

It should be interesting first to remember that the optical model (\ref{eq32})
with $\Delta^0_2 \rightarrow \Delta_{\rm CDW}$
gives a correct prediction for the $T \approx 0$~K
subgap conductivity characterizing  the ordered CDW systems, 
due to the gauge-invariance
factor $ (\hbar \omega)^2/E_{\underline{C}C} ^2 (  {\bf k})$ \cite{KupcicPB}.
	It also provides the correct treatment of 
the conductivity sum rules,
even for the relatively large damping energies, due to the factor 
$4 \hbar \omega E_{\underline{C}C} (  {\bf k})/[(\hbar  \omega  
+  E_{\underline{C}C} (  {\bf k}))^2  +  \hbar \Gamma_{\rm inter}^2]$,
at variance with the ordinary CDW model \cite{LRA,Gruner,Maki} which
fails in both of these cases.

   \begin{figure}[tb]
     \includegraphics[height=18pc,width=18pc]{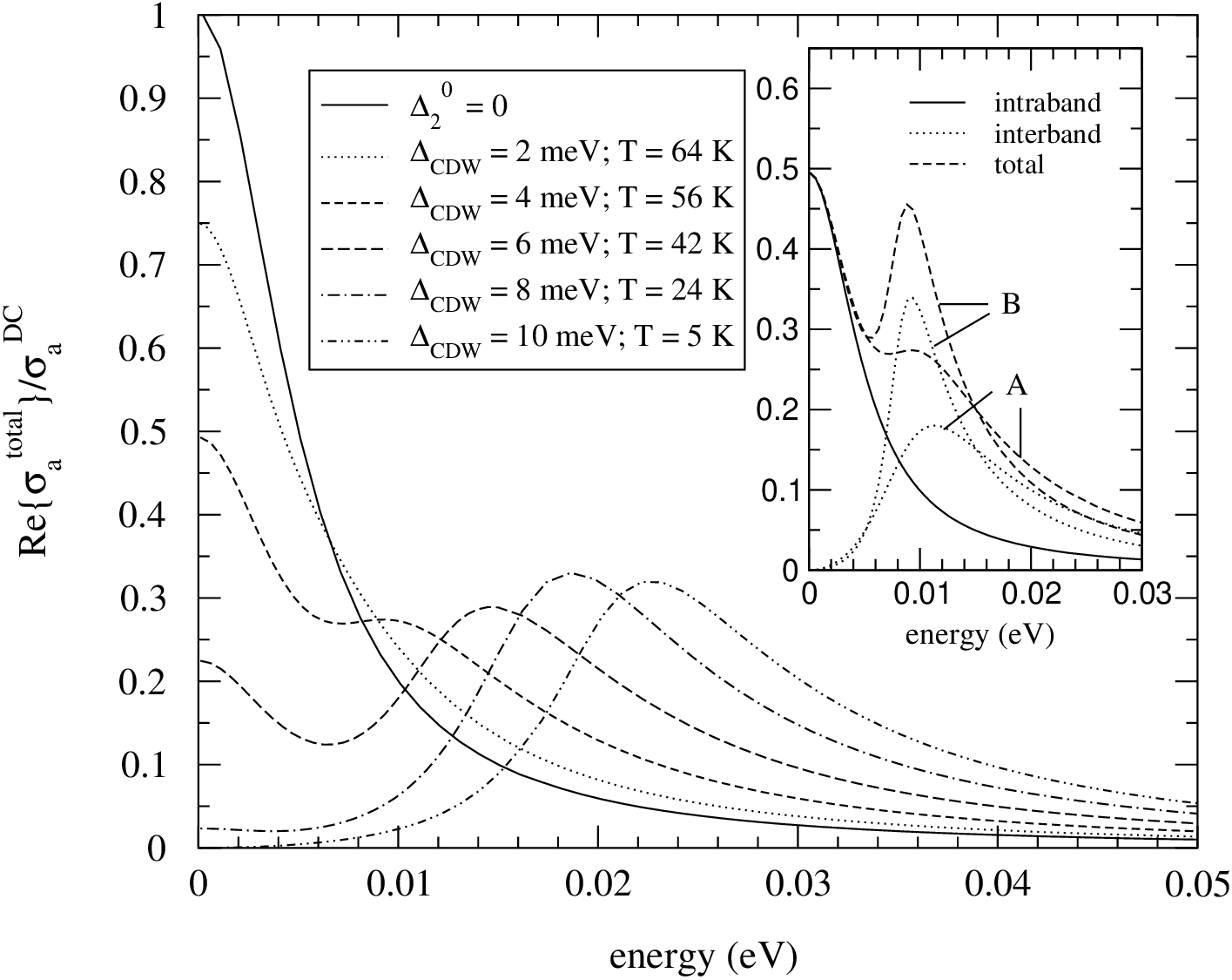}
    \caption{The  dependence of the real part of the  $\alpha = a$ 
    optical conductivity on $k_{\rm B}T$,  for $\delta = 1$, 
    $2\tilde{t}_{\rm a} = 0.25$ eV, 
    $t_{{\rm b}'} \rightarrow 0$ and $\hbar \Gamma_{\rm intra} = 5$ meV.
    Main figure:   The temperature dependence of the spectra for 
    $\Delta^0_2 \rightarrow \Delta_{\rm CDW} \le 10$ meV and  
    $\hbar \Gamma_{\rm inter} = 5$ meV (underdamped regime).
    $\sigma_{\rm a}^{\rm DC} $ is the DC conductivity of the metallic 
    $\Delta^0_2 = 0$ state, and the solid line is the related optical
    conductivity.  
    Inset of figure: The dependence of the total and interband spectra on
    $\hbar \Gamma_{\rm inter} $: 
    $\Delta_{\rm CDW} = 4$ meV, $T = 56$ K,
    $\hbar \Gamma_{\rm inter} = 5$ meV (curves A) and 
    $\hbar \Gamma_{\rm inter} = 2$ meV (curves B).  
    }
     \end{figure}

Fig. 4 illustrates the development of 
${\rm Re} \{ \sigma_{{\rm a}}^{\rm total} (\omega) \}$ with temperature in the 
CDW ground state for the underdamped case 
$\Delta_0/(\hbar \Gamma_{\rm inter}) = 2$.
	The temperature dependence of the order parameter is modeled by 
a simple expression $\Delta_{\rm CDW} (T) \approx \Delta_0  \sqrt{1 - T/T^*}$  
($T^*$ is an auxiliary temperature scale given by 
$k_{\rm B} T^* = \Delta_0  /1.75$, with $T^* \approx T_{\rm CDW}$).
	Interestingly, almost the same temperature behaviour is found for the 
$\alpha = b'$ optical conductivity in (TMTSF)$_2$PF$_6$ at 
$T < T^* \approx 15$~K, with the SDW 
(pseudo)gap parameter $2\Delta_0/(hc) \approx 70$~cm$^{-1}$ \cite{Degiorgi2}.

   \begin{figure}[tb]
     \includegraphics[height=18pc,width=18pc]{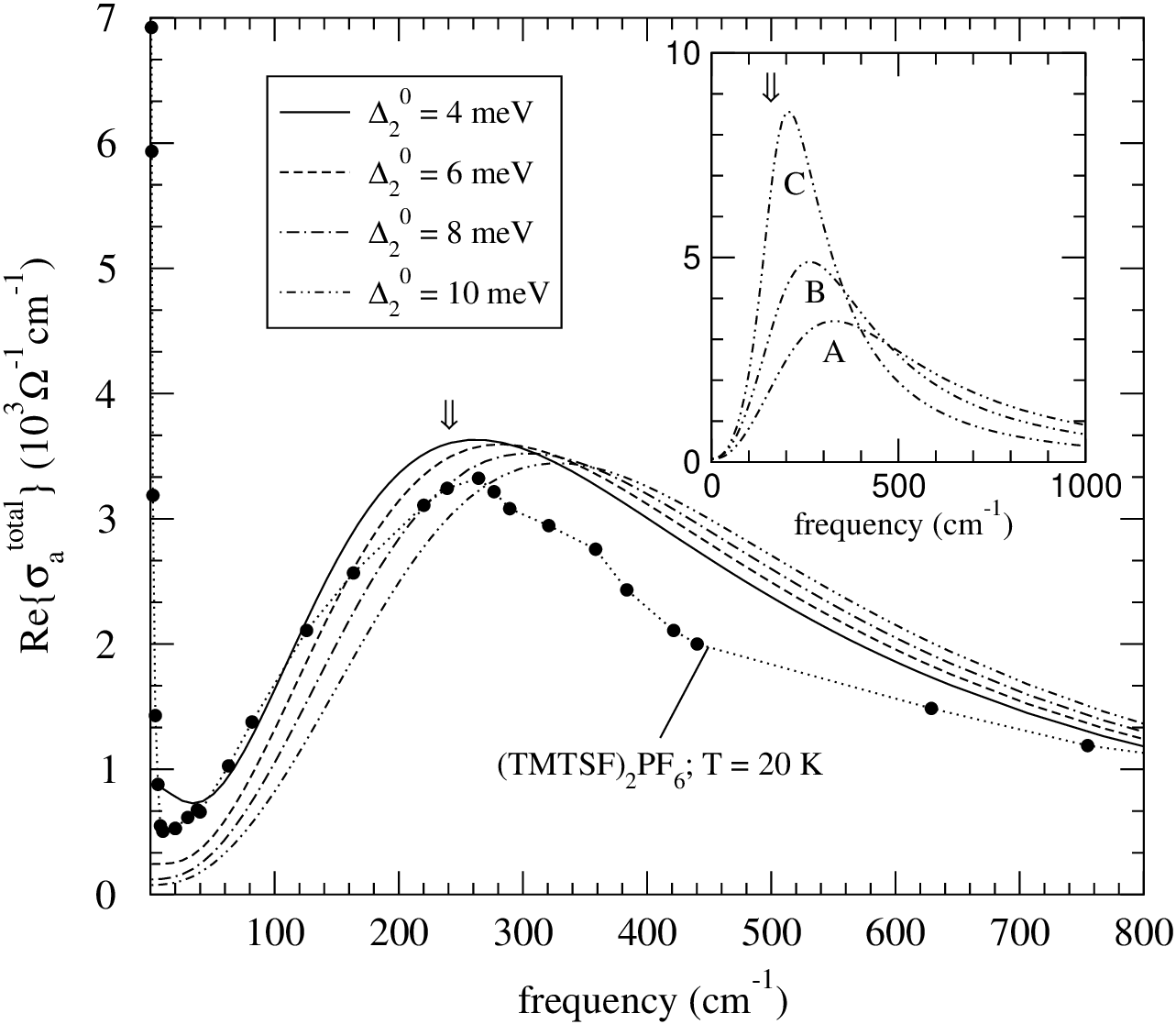}
    \caption{The  dependence of the real part of the  $\alpha = a$ 
    optical conductivity on the ratio $\hbar \Gamma_{\rm inter} / \Delta_2^0$,  
    for $\delta = 1$, $2\tilde{t}_{\rm a} = 0.25$ eV, 
    $t_{{\rm b}'} \rightarrow 0$,  $\hbar \Gamma_{\rm intra} = 5$ meV,
    $T = 10$ K, $m_{\rm aa} / m =1$ and 
    $\hbar \Omega_0^2/(4 \pi  \; {\rm eV}) = 0.355 \cdot 10^3 
    (\Omega {\rm cm})^{-1}$.
    Main figure:  The overdamped regime; $\hbar \Gamma_{\rm inter} = 30$ meV.
    The data measured in (TMTSF)$_2$PF$_6$  at $T = 20$ K are given 
    for comparison.
    Inset of figure: The intermediate regime; $\Delta_2^0 = 10$ meV with 
    $\hbar \Gamma_{\rm inter} = 30$, 20  and 10 meV for the curves A, B 
    and C, respectively. 
    The energies $\hbar \Gamma_{\rm inter}$ (main figure) and 
    $2\Delta_2^0$ (inset of figure) are indicated 
    by  arrows.
    }
     \end{figure}

Equally important is the observation in the normal metallic state of 
Bi$_2$Sr$_2$CuO$_6$ that the MIR maximum shifts with increasing 
temperature from 
$\omega_{\rm peak}/(2 \pi c) \approx 100$ cm$^{-1}$ at $T = 30$~K to 
$\omega_{\rm peak}/(2 \pi c) \approx 500$ cm$^{-1}$ at $T = 300$~K \cite{Lupi} 
(with a similar trend found in the La$_{2-x}$Sr$_x$CuO$_4$ 
compounds \cite{Venturini}).
	This unusual behaviour has been explained using the 1D orbital Kondo
model of Emery and Kivelson \cite{Kivelson}, which predicts 
$E_{\rm peak} \approx \hbar \Gamma$.
	Fig. 5 demonstrates that the same behaviour emerges
in the overdamped regime of the model (\ref{eq25})--(\ref{eq26}):
$\hbar \Gamma _{\rm inter} \gg \Delta^0_2$ leads to 
$E_{\rm peak} \approx \hbar \Gamma_{\rm inter} $.
	It seems that at least in the  compounds mentioned above the 
unusual temperature 
dependence of $E_{\rm peak}$ at  temperatures up to room temperature has the 
same origin, the competition between the correlation energy scale in question 
and the presumably large (interband) damping energy.

In conclusion, in the underdamped regime of the present 
optical model the position of the 
maximum in the MIR spectra is close to the threshold energy $E_{{\rm g}2}$ as well 
as to the correlation energy scale $2\Delta^0_2$.
	On the other hand, in the intermediate and overdamped regime 
one expects to be 
$E_{{\rm peak}} (\Delta^0_2, \hbar \Gamma_{\rm inter}) $, so that the 
correlation energy $2\Delta^0_2$ cannot be extracted directly from the measured
$E_{{\rm peak}}$.

\subsection{(TMTSF)$_2$PF$_6$}

Once the bond-energies $t_{\rm a}$ and $t_{{\rm b}'}$, together with the
correlation energy scale $2\Delta^0_2$, are estimated from the spectral weights,
the damping energies $\hbar \Gamma_{\rm intra}$ and $\hbar \Gamma_{\rm inter}$
and the  background dielectric function $\varepsilon_{\infty,{\rm a}}$ 
are the only free parameters to be estimated from the real part
of the optical conductivity and the real part of the dielectric function
obtained in  experiments.

According to the general relation 
\begin{eqnarray}
\varepsilon_{{\rm a}} (\omega) &\approx& \varepsilon_{\infty,{\rm a}} (\omega)
+ \frac{4 \pi \mathrm{i}}{\omega }   \sigma^{\rm total}_{{\rm a}} (\omega),
\label{eq33} \end{eqnarray} 
$\varepsilon_{\infty,{\rm a}} (\omega) - 1$  describes the contribution 
of the optical processes not included in the Hamiltonian (\ref{eq1}). 
	In the Bechgaard salts, for $\hbar \omega < 1$ eV, 
it is established that 
${\rm Im} \{\varepsilon_{\infty,{\rm a}} ( \omega) \}\approx 0$ and
${\rm Re} \{ \varepsilon_{\infty,{\rm a}} ( \omega) \} =
\varepsilon_{\infty,{\rm a}} \approx 2.5$
\cite{Bechgaard,Jacobsen}.

   \begin{figure}[tb]
     \includegraphics[height=18pc,width=18pc]{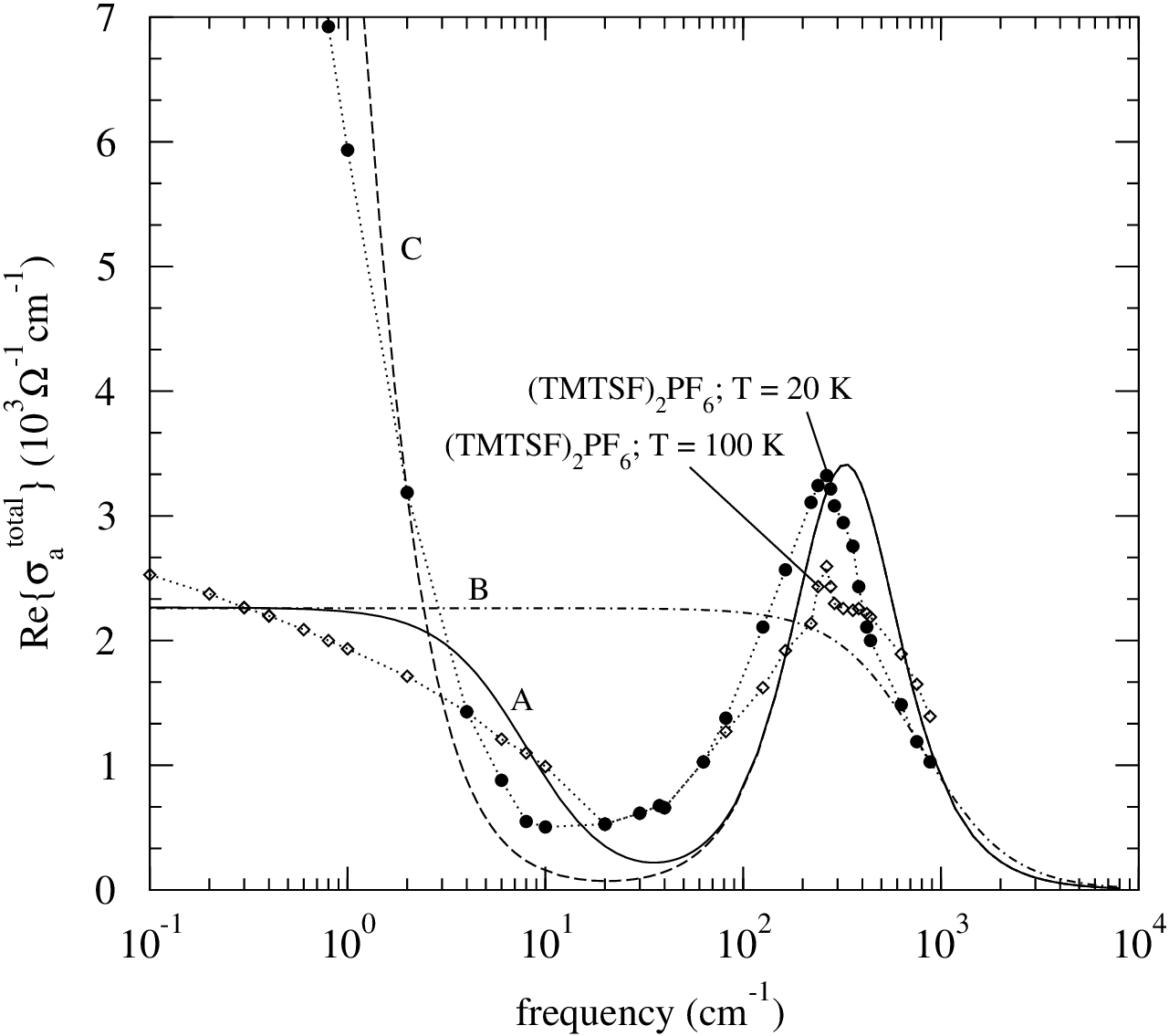}
    \caption{ 
    The real part of the $\alpha = a$ optical conductivity as 
    a function of the intraband damping energy,
    for the SD model with $n^{\rm eff}_{\rm intra, a} = n /100$;
    $2\tilde{t}_{\rm a} = 0.25$ eV, $2\tilde{t}_{{\rm b}'} = 8$ meV, 
    $\Delta^0_2  = 10$ meV, $\hbar \Gamma_{\rm inter} = 30$ meV,
    $T = 10$ K, $m_{\rm aa} / m =1$ and 
    $\hbar \Omega_0^2/(4 \pi  \; {\rm eV}) = 0.355 \cdot 10^3 
    (\Omega {\rm cm})^{-1}$.
    $\hbar \Gamma_{\rm intra} = 1$ and 0.1 meV for the curves A 
    and C, respectively.
    The prediction of the simple Drude model (curve B, $\Delta^0_2  = 0 $ and 
    $\hbar \Gamma_{\rm intra} = 0.1$ eV) 
    and the typical  measured spectra  \cite{Dressel}
    are also shown.
    }
     \end{figure}

In order to obtain a satisfactory explanation of the estimated damping energies
$\hbar \Gamma_{i}$, it is useful first to  remember
\cite{Gotze,Doniach,Abrikosov} that
the equation of motion (\ref{eq14}) generates automatically 
the self-energy and vertex contributions to 
the memory functions (i.e. to $\hbar \Gamma_{i}$).
	Furthermore, it should be noticed that omitting the vertex corrections 
makes the damping energies 
$\hbar \Gamma_{\rm intra}$ and $\hbar \Gamma_{\rm inter}$ appear as two limiting
cases of a frequency-dependent relaxation function 
$\hbar \Gamma (\hbar \omega)$ which describes the  pseudogap effects in the 
electron self-energy
($\hbar \Gamma_{\rm intra} \approx \hbar \Gamma ( 0)$ and
$\hbar \Gamma_{\rm inter} \approx \hbar \Gamma ( \hbar \omega \gg 2\Delta^0_2)$).
	Not surprisingly, the function $\hbar \Gamma (\hbar \omega)$ related 
to the umklapp scattering processes beyond the mean-field approximation 
adopted here should have the temperature 
and frequency dependence similar to that found  recently
for the half-filled spinless Holstein model in which the CDW (pseudo)gap
features have been studied  \cite{Perroni}.

   \begin{figure}[tb]
     \includegraphics[height=18pc,width=18pc]{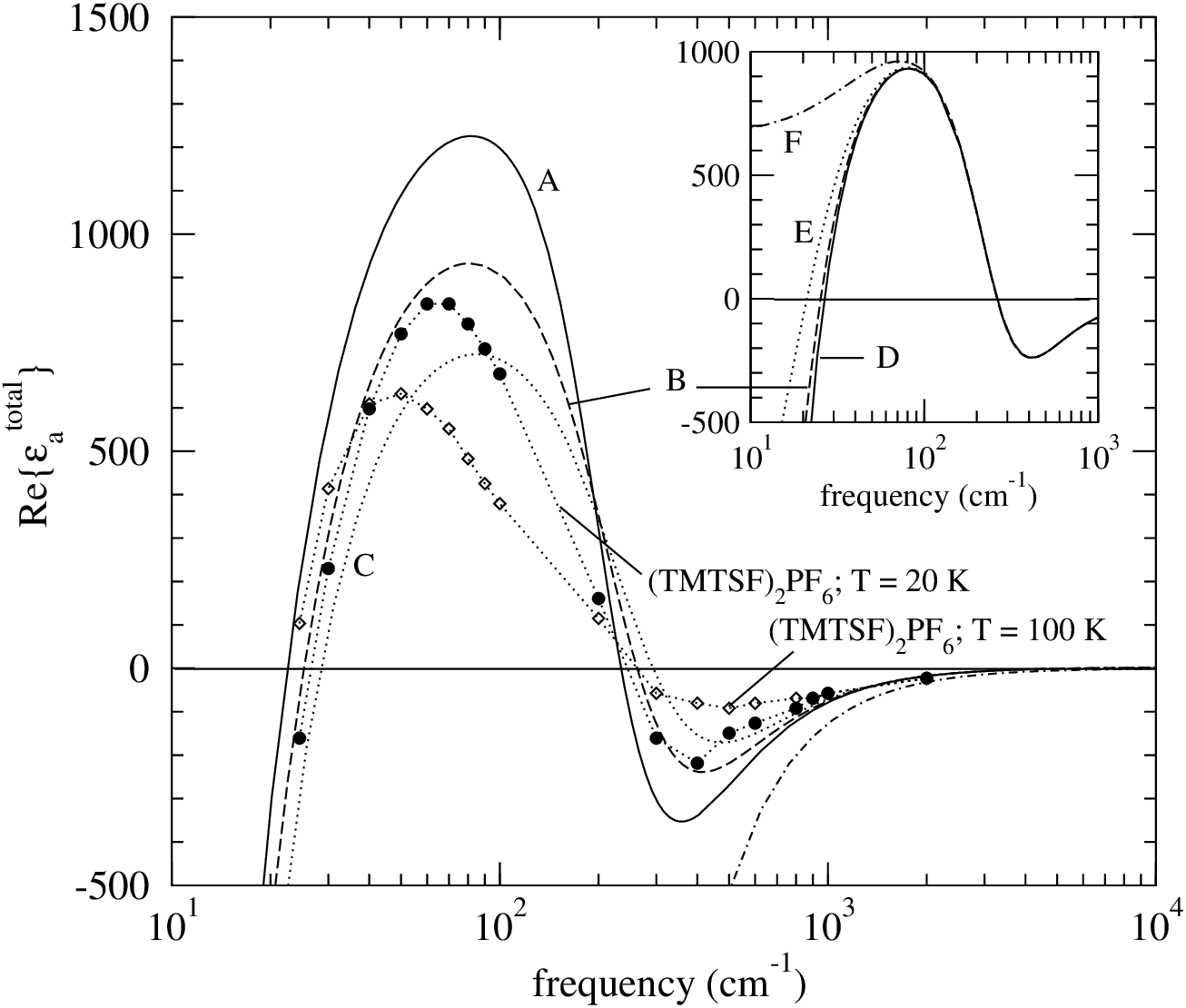}
    \caption{ 
    The dependence of the real part of the $\alpha = a$ dielectric function     
    on $\hbar \Gamma_{\rm inter}$ (main figure) and 
    $\hbar \Gamma_{\rm intra}$ (inset of figure), for $2\tilde{t}_a = 0.25$ eV,
    $2t_{{\rm b}'} = 8$ meV $\Delta^0_2  = 10$ meV, $T = 10$ K,
        $m_{aa}/m = 1$,
    $\hbar \Omega_0 = 1.4$ eV and $\varepsilon_{\infty,a} = 2.5$.
    Main figure: 
    $\hbar \Gamma_{\rm intra} = 1$ meV, and 
    $\hbar \Gamma_{\rm inter} = 15$, 20 and 25 meV for the curves A, B and C,
    respectively.
    The dot-dashed line describes the $\Delta_2^0 = 0$ case.
    Inset of figure:
     $\hbar \Gamma_{\rm inter} = 20$ meV and 
    $\hbar \Gamma_{\rm intra} = 0.1$,  1, 2 and 5 meV (D, B,  E and F).     
    The experimental data   are from Ref. \cite{Dressel}.
    }
     \end{figure}

The dependence of ${\rm Re} \{ \sigma^{\rm total}_{\rm a } (\omega ) \}$ and
${\rm Re} \{ \varepsilon^{\rm total}_{\rm a } (\omega ) \}$ 
on the damping energies $\hbar \Gamma_{i}$ for the SD case with 
$n^{\rm eff}_{\rm intra, a} = n /100$ is shown in Figs.~6 and 7, and compared
to the experimental data.
	According to the questions raised in Section 1, the most important
qualitative conclusions are as follows.

The power-law frequency dependence 
${\rm Re} \{ \sigma^{\rm total}_{\rm a } (\omega ) \} 
\propto \omega^{ - \nu} $, $\nu \approx 1.3$, 
for frequencies greater than $E_{\rm peak}/ \hbar$, observed  at
$T < 20$ K, is well understood as a direct consequence of the umklapp
scattering processes \cite{Giamarchi,Vescoli,Schwartz}. 
On the other hand, the temperature dependence of both 
$\nu $ and $E_{\rm peak}$, in particular at temperatures above 100 K,
indicates the regime where $\hbar \Gamma_{\rm inter}$ is comparable 
with $2\Delta_2^0$, and, consequently, where the temperature effects are 
primarily due to $\hbar \Gamma_{\rm inter} (T)$, as shown in the inset of 
Fig. 5.
	Therefore, the energy $\hbar \Gamma_{\rm inter} \approx$~25--30 meV 
represents roughly the high-frequency limit of the relaxation function 
$\hbar \Gamma (\hbar \omega)$.
	On the other hand, the zero-frequency limit reveals the energy 
$\hbar \Gamma_{\rm intra}  \approx 1$  meV at $T = 100$ K, indicating
the strong pseudogap effects, with an additional, pronounced decrease 
on decreasing temperature  ($\hbar \Gamma_{\rm intra}  \approx 0.1$  
meV at $T = 20$ K).
	To improve the overall agreement between the model predictions and the
measured spectra, including the anomalous frequency dependence of the Drude
peak as well as the  temperature dependence of its spectral weight, 
it is necessary to take into account the frequency dependence of the 
damping energies in Eqs. (\ref{eq25})--(\ref{eq26}).
	When considering  the two-component model with the causality properties
treated correctly, this problem,  however, requires a more accurate and 
extensive investigation and will not be discussed here.

According to Fig. 7, there are two zeros of 
${\rm Re} \{ \varepsilon_{\rm a} (\omega ) \}$,
representing the motion of two different plasma modes.
	These frequencies are related to the frequencies 
$\Omega_{i, a}$ estimated from the spectral weights.
	Not surprisingly,  a simple Drude relation 
$\omega^i_{{\rm pl}, {\rm a}} \approx \Omega_{i, {\rm a}} / 
\sqrt{\varepsilon_{\infty, {\rm a}}}$
is not valid here.
    	In this respect, the total plasma frequency 
$\omega^{\rm total}_{{\rm pl},  {\rm a}} /(2 \pi c)
 \approx 0.55 \cdot 10^4 \; {\rm cm}^{-1}$
has to be compared to the plasma frequency 
$\Omega_0 /(2 \pi c \sqrt{\varepsilon_{\infty,  {\rm a}}}) 
\approx 0.7 \cdot 10^4 \; {\rm cm}^{-1}$ of the $\Delta^0_2  = 0 $
case, and to the bare plasma frequency 
$\Omega_0 /(2 \pi c)  
\approx 1.1 \cdot 10^4 \; {\rm cm}^{-1}$
corresponding to $\Delta^0_2  = 0 $, $\varepsilon_{\infty,  {\rm a}} = 1$.
	Similarly, 
$\omega^{\rm intra}_{{\rm pl}, {\rm a}} /(2 \pi c)\approx 20 \; {\rm cm}^{-1}$, 
in contrast to 
$\Omega_{{\rm intra}, {\rm a}}/(2 \pi c) \approx 10^3 \; {\rm cm}^{-1}$.
	Furthermore, a pronounced dependence of 
$\omega^{\rm intra}_{{\rm pl}, {\rm a}} $ on $\hbar \Gamma_{\rm intra}$ 
for $\Gamma_{\rm intra} > \omega^{\rm intra}_{{\rm pl}, {\rm a}}$ 
can also be noticed in the inset of figure.

\section{Conclusion}
To summarize, we have formulated a gauge invariant
approach to the optical conductivity of the Q1D interacting
electronic systems with a MIR structure
clearly distinguished from both the zero-frequency (Drude-like) and
the high-frequency (background) contributions.
	By using the simplest limit of this model, 
with the frequency-independent damping energies, 
the confinement of the
conduction electrons to the highly conducting chains is analyzed.
 	The estimated ratio $2 \Delta_2^0/t_{{\rm b}'}$ required to give the 
spectral weights measured in (TMTSF)$_2$PF$_6$ agrees 
with the conclusions of the previous renormalization-group study.

The real part of the optical conductivity is calculated for the
typical insulating and metallic state, giving rise a satisfactory
agreement with the experimental findings at temperatures well above 
$T_{\rm SDW}$.
	The estimated damping energies clearly show the (pseudo)gap features
in the charge excitation spectrum and are in a qualitative agreement with the 
relaxation function calculated recently for the systems with the CDW ground
state.
	However, to obtain a better fit to the measured data, even at 
$T \gg T_{\rm SDW}$, one has to go beyond the present model and to treat 
the umklapp scattering processes dynamically.

Two plasma modes are found in the metallic state, 
with a nontrivial relation between their frequencies and  the 
corresponding spectral weights.

\section*{Acknowledgement}
This research was supported by the Croatian Ministry of Science and Technology 
under  the project 0119-256.

\appendix

\section{Vertex functions}
%

To account for the effects of the dimerization/correlation  potential on the
coupling between the tight-binding electronic system and the 
electromagnetic fields, we choose the $l{\bf k}$ representation, apply 
the usual substitution for the electron momentum, and, finally, 
transform the obtained coupling Hamiltonian in the $L{\bf k}$ representation.

Inserting
\begin{eqnarray}
\delta  H_0^{ll'} ({\bf k})  &\approx &  
- 
\frac{\partial  H_0^{ll'} ({\bf k})}{\partial k_{\alpha}}
\frac{e}{\hbar c} A_{\alpha} ({\bf q}) 
 \nonumber \\
 &&
+ \frac{1}{2} 
\frac{\partial^2 H_0^{ll'} ({\bf k})}{\partial k_{\alpha}^2}
\bigg( \frac{e}{\hbar c}\bigg)^2  A_{\alpha}^2  ({\bf q})  
\label{eqA1}
\end{eqnarray}
in
\begin{eqnarray}
  H^{\rm ext} &=& 
   \sum_{ll'{\bf k} \sigma}  \delta H_0^{ll'} ({\bf k}) 
  \big[ l_{{\bf k} + {\bf q} \sigma}^{\dagger} l'_{{\bf k} \sigma}
  + {\rm h.c.} \big],
\label{eqA2}
\end{eqnarray}
we obtain the  coupling Hamiltonian given by Eq. (\ref{eq9}) in the main text.
	The  intraband bare Raman vertices  are of the form
\begin{eqnarray}
&&\gamma ^{\underline{C}\underline{C},CC}_{\alpha \alpha} ( {\bf k};2)  = 
  \frac{m}{2\hbar^2} 
\bigg[  
\frac{\partial^2 [\varepsilon_{\underline{c}} ({\bf k}) 
+ \varepsilon_{c} ({\bf k})]}{ \partial k_{\alpha}^2}  
\nonumber \\ 
&&
\hspace{5mm} \pm \cos \varphi ({\bf k})  
\frac{\partial^2 \varepsilon_{\underline{c}c} ({\bf k})}{ \partial k_{\alpha}^2} 
\pm \sin \varphi ( {\bf k}) 
\frac{2 \partial^2 |\Delta ({\bf k})|  }{\partial k_{\alpha}^2}\bigg],
\label{eqA3}\end{eqnarray}
with the upper (lower)  sign in $\pm$ corresponding to 
$\underline{C}\underline{C}$ ($CC$).
	Similarly,  the intra- and interband current vertices are 
\begin{eqnarray}
&&J_{\alpha} ^{\underline{C}\underline{C},CC} ( {\bf k})  =
  \frac{e}{2\hbar} 
\bigg[  
\frac{\partial [\varepsilon_{\underline{c}} ({\bf k}) 
+ \varepsilon_{c} ({\bf k})]
}{ \partial k_{\alpha}}  
 \nonumber \\ 
 &&
\hspace{5mm} \pm \cos \varphi ({\bf k})  
\frac{\partial \varepsilon_{\underline{c}c} ({\bf k})}{ \partial k_{\alpha}} 
\pm \sin \varphi ( {\bf k}) 
\frac{2 \partial |\Delta ({\bf k})|  }{\partial k_{\alpha}}\bigg],
 \label{eqA4} 
  \\ 
&&J_{\alpha} ^{\underline{C}C} ( {\bf k})  = 
\big[ J_{\alpha} ^{C\underline{C}} ( {\bf k}) \big]^* = 
  -\frac{e}{2\hbar}  e^{-\mathrm{i} \phi_0} 
 \nonumber \\ 
 && 
 \hspace{5mm} \times \bigg[
 \sin \varphi ({\bf k})  
\frac{\partial \varepsilon_{\underline{c}c} ({\bf k})}{ \partial k_{\alpha}} 
- \cos \varphi ( {\bf k}) 
\frac{2 \partial |\Delta ({\bf k})|  }{\partial k_{\alpha}}\bigg].
\label{eqA5}
 \end{eqnarray}
	The corresponding static Raman vertices 
$\gamma^{\underline{C}\underline{C},CC}_{\alpha \alpha} ( {\bf k})$
come from the effective mass theorem
\begin{eqnarray}  
 \gamma^{\underline{C}\underline{C},CC}_{\alpha \alpha} ( {\bf k}) 
 &=&  \gamma^{\underline{C}\underline{C},CC}_{\alpha \alpha} ( {\bf k};2) 
\pm \frac{2m|J^{\underline{C}C}_{\alpha} ( {\bf k})|^2}{  
e^2E_{\underline{C}C} ( {\bf k})} \nonumber \\
&\equiv& \frac{m}{\hbar^2} 
\frac{\partial^2 E_{\underline{C},C} ({\bf k})}{ \partial k_{\alpha}^2}.
\label{eqA6}
 \end{eqnarray}

It is interesting also to note that for the case considered in Section~3 
($\varepsilon_{\underline{c}} ({\bf k}) = - \varepsilon_{c} ({\bf k})$, 
$\Delta ({\bf k}) = \Delta^0_2$), the $\alpha = a$ vertices read as
\begin{eqnarray}
\gamma ^{\underline{C}\underline{C},CC}_{\rm aa} ( {\bf k};2)  &=& 
\mp \cos \varphi ({\bf k})  \gamma ^{cc}_{\rm aa} ( {\bf k};2), 
\nonumber \\
J_{\rm a} ^{\underline{C}\underline{C},CC} ( {\bf k})  &=&  
  \mp\cos \varphi ({\bf k})  J_{\rm a} ^{cc}( {\bf k}) , 
\nonumber \\ 
J_{\rm a} ^{\underline{C}C} ( {\bf k})  &=&    
\sin \varphi ({\bf k})  J_{\rm a} ^{cc} ( {\bf k}) ,
\label{eqA7}
 \end{eqnarray}
where 
$\gamma ^{cc}_{\rm aa} ( {\bf k};2) = 
\gamma ^{cc}_{\rm aa} ( {\bf k}) = (m/\hbar^2) 
\partial^2 \varepsilon_{c} ({\bf k})/ \partial k_{\rm a}^2$
and 
$J_{\rm a} ^{cc} ( {\bf k})  = (e/\hbar) \; 
\partial \varepsilon_{c} ({\bf k})/ \partial k_{\rm a}$
are the vertices of the $\Delta^0_2 = 0$ model.


\begin{references}
\bibitem{comment1}
Notice that the approximation $n_{{\rm \rm intra}, \alpha}^{\rm eff} \approx n$ 
is appropriate when the carriers behave as ordinary electrons/holes, including 
the tight-binding electron/hole systems below quarter-filling; here $n$ is 
the electron/hole concentration.
\bibitem{comment4}
Notice the maxima in $n_{{\rm total,b}'}^{\rm eff}$ which separate
the doping range corresponding to the open Fermi surface from the ones
representing the closed Fermi surfaces.
\end{references}
\end{document}